\documentclass[10pt,letterpaper]{article}
\usepackage[top=0.85in,left=2.75in,footskip=0.75in]{geometry}
\usepackage{amsmath,amssymb}
\usepackage{changepage}
\usepackage[utf8]{inputenc}
\usepackage{textcomp,marvosym}
\usepackage{cite}
\usepackage{nameref,hyperref}
\usepackage[right]{lineno}
\usepackage{microtype}
\DisableLigatures[f]{encoding = *, family = * }
\usepackage[table]{xcolor}
\usepackage{array}
\usepackage{siunitx}                 
\newcolumntype{+}{!{\vrule width 2pt}}
\newlength\savedwidth

\raggedright
\usepackage[aboveskip=1pt,labelfont=bf,labelsep=period,justification=raggedright,singlelinecheck=off]{caption}

\makeatletter
\renewcommand{\@biblabel}[1]{\quad#1.}
\makeatother
\usepackage{lastpage,fancyhdr,graphicx}
\usepackage{epstopdf}
\fancyheadoffset[L]{2.25in}
\fancyfootoffset[L]{2.25in}
\newif\ifsubmission
\submissionfalse
\let\realincludegraphics\includegraphics
\renewcommand{\includegraphics}[2][]{%
  \ifsubmission
    \textit{[Figure uploaded separately; see figure files.]}%
  \else
    \realincludegraphics[#1]{#2}%
  \fi}

\begin{document}
\vspace*{0.2in}

\begin{flushleft}
{\Large
\textbf\newline{Quantifying reticulocyte biomechanics in health and disease}
}
\newline
\\
Zhaojie Chai\textsuperscript{1,\Yinyang*},
Jianlu Zheng\textsuperscript{2,\Yinyang},
He Li\textsuperscript{3},
Ming Dao\textsuperscript{2*},
George Em Karniadakis\textsuperscript{1*}
\\
\bigskip
\textbf{1} Division of Applied Mathematics, Brown University, Providence, Rhode Island, United States of America
\\
\textbf{2} Department of Materials Science and Engineering, Massachusetts Institute of Technology, Cambridge, Massachusetts, United States of America
\\
\textbf{3} College of Engineering, University of Georgia, Athens, Georgia, United States of America
\\
\bigskip

\Yinyang These authors contributed equally to this work.

* zhaojie\_chai@brown.edu (ZC); mingdao@mit.edu (MD); george\_karniadakis@brown.edu (GEK)

\end{flushleft}

\section*{Abstract}
Red blood cell (RBC) populations are mechanically and morphologically heterogeneous, yet how this heterogeneity governs transport, splenic retention and blood rheology in confinement remains unclear. We combine microchannel experiments with dissipative particle dynamics (DPD) simulations to determine how the morphology, deformability and hydrodynamic interaction of immature RBCs (reticulocytes) shape microconfined flow in acute and chronic mountain sickness. Reticulocyte-rich samples present the three shapes that the maturation literature assigns to successive reticulocyte stages---multilobular, cup-shaped and near-discocytic---which we model as R1--R3, each with a membrane shear modulus within the range reported for reticulocytes. A $5$-\si{\micro\metre} microchannel resolves these models only weakly, R1 taking at most $\sim\!18\%$ longer than control discocytes to cross it, whereas a $1.5$-\si{\micro\metre} splenic slit, driven at its own loading, delays R1 by $\sim\!30\%$; both geometries order the models by shear modulus. In cell pairs, a leading cell never lets a follower pass below its own single-cell threshold, ruling out the order-of-magnitude reduction that a wake-``unjamming'' picture would suggest; instead the follower queues behind the leader, and a stiffer leader, which clears the slit more slowly, delays it more. The controlling variable is the single-cell critical pressure gradient $\Delta P_c$, which rises monotonically from control discocytes through the reticulocyte subtypes to sickled cells. Set against published rheograms, an analytical estimate attributes chronic-mountain-sickness hyperviscosity mainly to hematocrit-driven crowding rather than single-cell rheology. These results place benign acclimatization, chronic-mountain-sickness hyperviscosity and sickle-cell-trait splenic syndrome on a single mechanical axis defined by $\Delta P_c$.

\section*{Author summary}
Our bodies must keep red blood cells flowing through passages far narrower than the cells themselves---nowhere more so than in the spleen, whose slit-like filters clear aged or damaged cells. Young red cells (reticulocytes), released in large numbers during stresses such as the sustained low oxygen of high altitude, are larger and more variable in shape than mature cells and span a wide range of deformability, and how this diversity affects their passage has been unclear. Combining microchannel experiments with particle-based simulations, we show that whether an individual cell squeezes through a splenic-slit-scale pore is governed by a single mechanical quantity, the critical pressure gradient, which rises steadily as cells become less deformable, from mature discocytes through the reticulocyte subtypes to sickled cells. We further find that a leading cell does not carry the cell behind it through a pore it could not pass alone; the follower simply waits its turn, so a soft leader, which clears the pore sooner, holds it up less than a stiff one. Placing these mechanics on a common axis, we explain the acute and chronic blood responses to altitude and clarify why sickle-cell-trait carriers can suffer splenic injury upon rapid ascent.


\section*{Introduction}

Red blood cells (RBCs) rely on extreme deformability to sustain oxygen transport through the microcirculation and to survive repeated passage through the spleen, where inter-endothelial slits (IES) impose some of the tightest mechanical constraints encountered in vivo~\cite{Mebius2005Structure,Cesta2006Normal,Li2018Mechanics,moreau2023physical}. A mature human RBC is an $\sim$8-\si{\micro\metre}-diameter biconcave disc that must traverse 2--3~\si{\micro\metre} capillaries and $\sim$1-\si{\micro\metre} splenic slits, a feat made possible by a delicate interplay of membrane shear elasticity ($\mu \sim 5$~\si{\micro\newton\per\metre}), bending rigidity ($k_c \sim 2 \times 10^{-19}$~J), cytoplasmic viscosity, and excess surface area~\cite{evans1973new,dao2021erythrocyte}. A typical cell in this state carries an area of $\sim\!135~\si{\micro\metre\squared}$ enclosing a volume of $\sim\!90$--$95$~fL, producing a surface-to-volume ratio $S/V \approx 1.44~\si{\per\micro\metre}$ that sets the deformation reserve available for geometric confinement~\cite{evans1973new,mohandas2008red}. Disrupting any of these properties---whether through hereditary disorders such as sickle cell trait, acquired states such as diabetes or Gaucher disease, or physiological stress such as hypoxia---alters microvascular transport, promotes mechanical retention in the spleen, and shifts bulk blood rheology~\cite{baskurt2003blood,nader2019blood,perazzo2022effect}. Understanding which mechanical descriptors control which flow regime is therefore central to microcirculatory biophysics.

Circulating blood is not a monodisperse suspension of mature cells: $0.5$--$2.5\%$ of RBCs in healthy adults are reticulocytes, and this fraction can rise severalfold during stress erythropoiesis induced by bleeding, hemolytic anemia, or sustained hypoxia~\cite{chasis1989membrane,clark1988senescence}. Reticulocytes released from the bone marrow are $\sim\!15$--$30\%$ larger in membrane area than their mature counterparts, retain residual organelles (ribosomal remnants, mitochondria, autophagic vesicles), and carry a spectrin--actin cytoskeleton that is only partially remodeled~\cite{malleret2013significant,mohandas2008red,Renoux2019Impact}. The resulting cell is morphologically heterogeneous, ranging from multilobulated shapes with several surface protrusions to cup-shaped intermediates and near-discocytes~\cite{mel1977reticulocyte,malleret2013significant,chasis1989membrane,gronowicz1984maturation,StevensHernandez2022Reticulocyte}; sorting human reticulocytes on RNA content and transferrin-receptor (CD71) expression has recently tied the multilobulated shape to the least mature of four fractions and the stomatocytic, cup-like shape to the two most mature~\cite{Leduc2025Comprehensive}. During a $24$--$48$~h maturation window, the reticulocyte progressively sheds excess membrane through vesiculation and pitting and reconfigures its cytoskeleton to reach the mature-discocyte state~\cite{pivkin2016biomechanics,moreau2023physical,li2021How}. Critically, these remodeling events are not purely cell-autonomous: they are spatially coupled to splenic passage and geometrically tuned to the IES, so that the circulating deformability envelope is set, in part, by where and how rapidly reticulocytes encounter the splenic filter.

The human spleen performs this filtration through a specialized vascular geometry. Red cells exit the arterioles into the red-pulp reticular meshwork, where they must squeeze through $0.5$--$1.5~\si{\micro\metre}$-wide, $2$--$5~\si{\micro\metre}$-tall inter-endothelial slits to re-enter the venous sinusoids~\cite{Mebius2005Structure,Cesta2006Normal,macdonald1987kinetics}; models of this passage place the associated trans-slit pressure drop at $\sim\!200$--$500$~Pa~\cite{dao2021erythrocyte,moreau2023physical}. Cells whose combined shear/bending deformability fails to meet the slit geometry at this pressure are retained in the cords and eventually phagocytosed~\cite{safeukui2012quantitative,safeukui2018sensing,chai2026multiscale}. The splenic filter is therefore not a simple size cutoff but a mechanical envelope defined jointly by $\mu$, $k_c$, and $S/V$. Reticulocyte mechanics sit near the edge of this envelope: their excess area offers abundant deformation reserve, but the partially remodeled cytoskeleton and residual organelles can locally raise the mechanical cost of IES entry. Whether a given reticulocyte traverses the IES freely, slows, or is retained---and whether neighboring cells are thereby helped or hindered---is set by the interaction between single-cell mechanics and local slit hydrodynamics.

High-altitude exposure provides a physiological context in which reticulocyte mechanics become central. Acute hypoxic stress ($\lesssim\!24$~h at $\ge\!2500$~m) triggers sympathetically driven splenic contraction, which rapidly expels a sequestered RBC reservoir and transiently raises the circulating RBC count by $\sim\!2$--$10\%$; this partially compensates for the reduced arterial oxygen saturation~\cite{stewart2002spleen}, while the hypoxia itself can precipitate the headache, gastrointestinal upset, fatigue and dizziness that define acute mountain sickness (AMS)~\cite{roach2018lakelouise}. Prolonged residence at altitude ($\ge\!$weeks) drives erythropoietin-mediated erythropoiesis, persistent reticulocytosis, and hematocrit elevations to $60$--$70\%$ in chronic mountain sickness (CMS), often accompanied by a $3$--$7\times$ rise in low-shear blood viscosity~\cite{stauffer2024making}. At the pathological limit, sickle-cell-trait (SCT) carriers, who are heterozygous (HbAS, with only $\sim\!40\%$ HbS versus $>\!80\%$ in sickle cell disease) and clinically silent under normal oxygenation, can, under the profound hypoxia of high altitude, undergo partial deoxy-HbS polymerization in a subpopulation of cells; the resulting moderate stiffening, acting within the low-oxygen environment of the splenic red pulp, can mechanically retain these cells and precipitate the acute ``splenic syndrome'' of mountaineers~\cite{goodman2014splenic}. The three phenotypes---benign AMS acclimatization, CMS hyperviscosity, and SCT splenic syndrome---thus sit on the same clinical axis but arise in populations whose circulating cells differ in ways that no single mechanical descriptor can summarize.

Existing studies of splenic filtration and microvascular transport have mapped single-cell response to geometry and pressure with high fidelity~\cite{quinn2011combined,Li2018Mechanics,dao2021erythrocyte,qiang2023microfluidic}, but have generally treated the RBC population as homogeneous. Blood flow through confined geometries, however, is inherently collective: local pressure barriers, cell--cell hydrodynamic interaction, and transient jamming all emerge from many-body interactions and cannot be predicted from isolated-cell behavior alone~\cite{perazzo2022effect,recktenwald2024morphology}. Reticulocyte heterogeneity introduces a further complication because compliant and stiff cells can interact through the fluid and through single-file crowding as they line up at a constriction: a leading cell may either ease the passage of the next cell in line or, if it lodges in the pore, obstruct it. Which of these outcomes dominates for splenic filtration and how it scales with reticulocyte fraction remains an open question, and one with direct implications for how the AMS, CMS, and SCT syndromes should be interpreted mechanically.

Motivated by this gap, and enabled by multiscale particle-based modeling~\cite{tang2017openrbc,chai2023dynamics,chai2022periodic,chang2016md,zhang2021deep,zhang2020deep}, we combine microfluidic-style experiments with dissipative-particle-dynamics (DPD) simulations~\cite{fedosov2010multiscale,Fedosov2011Multiscale,Fedosov2011Predicting,groot1997dissipative,hoogerbrugge1992simulating} to pursue four related goals: (i) to identify the mechanical and morphological descriptors that distinguish reticulocyte subtypes and to represent them in a single DPD model; (ii) to determine how these descriptors control single-cell transit in capillary-like channels and in splenic slits, each driven at the loading appropriate to it; (iii) to test whether cell--cell interaction in tandem transits facilitates or instead obstructs the passage of stiff cells through extreme constrictions; and (iv) to connect the single-cell passage thresholds and shear-dependent viscosity to the phenotypes of AMS, CMS, and SCT-associated splenic syndrome. We address these goals in turn and, by resolving single-cell and pairwise transport with a single simulation platform and reading them against suspension rheology, place clinically divergent altitude syndromes on a shared mechanical axis rather than treating them as unrelated phenomena.

\section*{Results}

The results that follow are organized around a single biophysical objective: \emph{to map how the mechanical properties of a young, not-yet-fully-remodeled red cell translate, via single-cell and collective hydrodynamics, into the hematological phenotypes that distinguish high-altitude acclimatization, chronic mountain sickness, and sickle-cell-trait splenic syndrome.} The central idea of this study is that clinically divergent altitude syndromes emerge as distinct operating points on a shared mechanical axis---namely, the critical pressure gradient $\Delta P_c$ required for IES passage, read relative to the control-discocyte threshold---computed for single cells and cell pairs on one DPD platform. To build this axis, we proceed from the cell up. We first establish a mechanical representation of three coexisting reticulocyte subtypes (R1--R3), assigning the constitutive parameters $(\mu, S/V, k_c)$ within independently reported ranges (Figs~\ref{fig:morphology}--\ref{fig:shape_transition}). We then probe how these cells transit physiologically relevant confined geometries (Fig~\ref{fig:microchannel}), and find that stiffness delays a reticulocyte nearly twice as much in the narrow splenic slit as in the wider microchannel. The pairwise interaction analyses (Fig~\ref{fig:pairwise}) test whether tandem interaction eases stiff-cell passage, and show that a leading cell never lets a follower pass below its own single-cell threshold: the follower queues behind the leader, whose compliance sets how long it waits, so that its passage threshold is $\sim\!5\%$ higher behind a stiff leader than behind a compliant one. We then quantify, for each cell population, the critical pressure gradient that separates slit passage from retention (Fig~\ref{fig:dpc_ladder}), and set these beside suspension-scale viscosity from published rheograms and from our earlier simulations on the same platform~\cite{chai2025silico} (Fig~\ref{fig:viscosity}). A concluding framework (Fig~\ref{fig:conceptual}) integrates all of these observables into a compact mechanistic map in which AMS, CMS, and SCT splenic syndrome occupy well-separated positions on the same $\Delta P_c$ axis, modulated by population crowding at elevated hematocrit.

\subsection*{Morphological heterogeneity and mechanical representation of reticulocyte models}

\begin{figure}[!tbp]
\begin{center}
\includegraphics[width=1.000\textwidth]{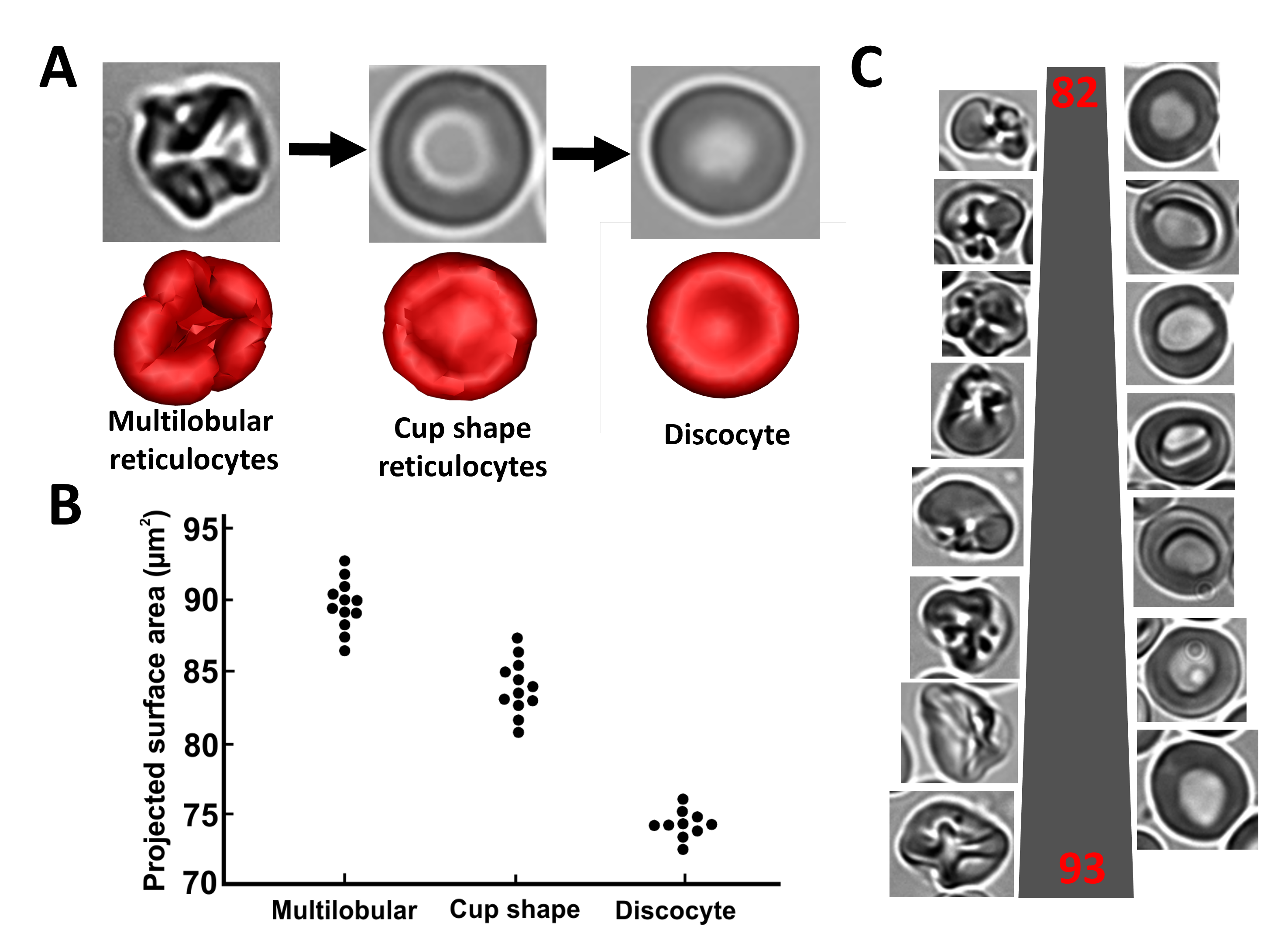}
\end{center}
\vspace{-0.15in}
\caption{\small{\bf Morphological subclasses and projected surface area of reticulocytes.}
(A) Representative bright-field images (top) and the corresponding DPD membrane configurations (bottom) for three major morphological subclasses observed in reticulocyte-rich blood samples: multilobular, cup-shaped, and near-discocyte. Multilobular cells exhibit irregular, fragmented surfaces with multiple lobes; cup-shaped cells display a unilobular concave depression; and discocytes retain a near-normal biconcave morphology. The lower row shows triangulated model meshes ($500$ vertices, $996$ faces) relaxed under a reduced enclosed volume at fixed membrane area (Methods); they are shape models used to connect a two-dimensional silhouette to a membrane area, not tomographic reconstructions of the cells imaged above.
(B) Distribution of projected surface areas for each morphological subclass. Multilobular reticulocytes show the largest projected areas ($\sim\!86$--$93~\si{\micro\metre\squared}$), followed by cup-shaped cells ($81$--$87~\si{\micro\metre\squared}$), while discocytes exhibit the smallest areas ($73$--$76~\si{\micro\metre\squared}$), reflecting the progressive reduction in surface area with reticulocyte maturation.
(C) Two representative columns of cells, multilobular (left) and cup-shaped (right), flanking a wedge that ranks them by increasing projected surface area, from $82~\si{\micro\metre\squared}$ (top) to $93~\si{\micro\metre\squared}$ (bottom), highlighting the morphological heterogeneity of cells in reticulocyte-rich samples and the coexistence of multiple subclasses within a single donor sample. This panel is illustrative of the morphological continuum rather than a quantitative ranking.}
\label{fig:morphology}
\end{figure}

Bright-field imaging of reticulocyte-rich blood samples resolves three predominant red-cell shape classes (Fig~\ref{fig:morphology}A), each of which we represent by a matched DPD membrane configuration: (i) highly irregular multilobulated reticulocytes with multiple surface lobes, (ii) unilobular cup-shaped cells with a single deep concavity, and (iii) near-biconcave discocytes that are morphologically indistinguishable from mature RBCs~\cite{mel1977reticulocyte,chasis1989membrane,malleret2013significant,cluitmans2012red,mohandas2008red,gronowicz1984maturation,chai2026quantifying}. These are the shapes that the maturation literature assigns to successive reticulocyte stages. The youngest reticulocytes are motile and multilobulated~\cite{mel1977reticulocyte,chasis1989membrane,StevensHernandez2022Reticulocyte}; as they mature they become ``deep dished''~\cite{mel1977reticulocyte}, ``deeply cupped''~\cite{gronowicz1984maturation}, or cup-shaped and nonmotile~\cite{chasis1989membrane}---a single-concavity form also visible in scanning electron micrographs of R2 reticulocytes~\cite{StevensHernandez2022Reticulocyte}---before flattening into the discocyte; and in reticulocytes sorted from human bone marrow on RNA content and CD71 expression, the multilobulated shape dominates the least mature of four fractions while the stomatocytic, cup-like shape characterizes the two most mature~\cite{Leduc2025Comprehensive} (those fractions, R1--R4 in that study, are numbered independently of our models; its R3 and R4 correspond to the cup-shaped stage that we model as R2). The stage assigned to each class, and hence the mapping onto R1--R3 below, rests on that literature: bright-field shape alone does not establish that a given cup-shaped or discoid cell is a reticulocyte, which we test directly by supravital staining below. Quantitative image analysis (Fig~\ref{fig:morphology}B) shows that projected surface area decreases monotonically with apparent maturation state: multilobular cells span $\sim\!86$--$93~\si{\micro\metre\squared}$, cup-shaped cells $81$--$87~\si{\micro\metre\squared}$, and discocytes $73$--$76~\si{\micro\metre\squared}$---a $17\%$ drop between the multilobular and discocyte midpoints, and $21\%$ between their extremes---and the ranked image panel (Fig~\ref{fig:morphology}C) illustrates a continuous $13\%$ spread from $82$ to $93~\si{\micro\metre\squared}$ across the multilobular and cup-shaped ranges, indicating that the three classes are best interpreted as coarse-grained bins of an underlying continuum. The reference volumes assigned to the models follow the same ordering, from $\sim\!103$~fL for the multilobular model (R1) to $\sim\!92$~fL for the mature discocyte (total span $\sim\!11\%$), consistent with the excess membrane and volume reserves that are progressively lost during membrane remodeling~\cite{chasis1989membrane,Renoux2019Impact}. The surface-to-volume ratios of the three models, $S/V \approx 1.56$, $1.52$ and $1.48~\si{\per\micro\metre}$, follow from the reference areas and volumes of Table~\ref{tab:mechanical_properties}, which are set from reported vesiculation and pitting losses relative to the mature-discocyte reference (Methods) rather than from the images; the imaging above enters as an independent morphological constraint on which shape each model should take. The $S/V$ separation is small. The shape-transition simulations below (Fig~\ref{fig:shape_transition}) carry it, whereas the transport simulations that follow them hold every model at the control value; restoring the listed $S/V$ there slightly shortens the reticulocyte delays at the slit and lowers the R1 threshold (\nameref{S1_Text}, Sections~5 and~7).

\begin{table}[htbp]
\centering
\caption{Mechanical properties for the control discocyte (CTR-RBC) and three reticulocyte models (R1--R3).}
\label{tab:mechanical_properties}
\resizebox{\textwidth}{!}{%
\begin{tabular}{lcccc}
\hline
 & \textbf{CTR-RBC} & \textbf{R1} & \textbf{R2} & \textbf{R3} \\
\hline
Shear modulus $\mu$ (\(\mu\mathrm{N/m}\)) & 4.73 & 8.28 & 7.00 & 6.29 \\
Surface-to-volume ratio $S/V$ (\si{\per\micro\metre}) & 1.44 (132.9/92.5) & 1.56 (160.0/102.6) & 1.52 (150.0/98.7) & 1.48 (142.0/95.9) \\
Bending modulus $k_{c}$ ($10^{-19}$~J) & 2.4 & 4.8 & 4.8 & 7.2 \\
\hline
\end{tabular}}
\begin{minipage}{\textwidth}
\vspace{0.1cm}
\small Notes: R1, multilobular reticulocyte; R2, cup-shaped reticulocyte; R3, near-discocytic (thick-disc) reticulocyte. $S/V$ entries give the ratio (surface area in \si{\micro\metre\squared} / volume in fL). The shear modulus is assigned within the range reported for reticulocytes; $A_0$ and $V_0$, and hence $S/V$, are the control-discocyte reference scaled by reported vesiculation and pitting losses; and $k_c$ is taken from the cited literature (Methods). In the shape-transition simulations (Fig~\ref{fig:shape_transition}) each reticulocyte model carries all three listed values, with its $S/V$ imposed by lowering the reference volume at the control membrane area ($A_0 = 132.9~\si{\micro\metre\squared}$; $V_0 = 85.4$, $87.6$ and $90.0$~fL for R1, R2 and R3). In the transport simulations that follow, all populations share the reference area, reference volume and bending modulus of the control discocyte ($A_0 = 132.9~\si{\micro\metre\squared}$, $V_0 = 92.5$~fL), so that they differ from one another in shear modulus alone; there the $S/V$ and $k_c$ values listed here enter only the tests of \nameref{S1_Text}, Sections~5 and~7.
\end{minipage}
\end{table}

This shape sequence reflects the ordered biochemical program of reticulocyte maturation. Young reticulocytes released from the bone marrow carry a substantial excess of membrane lipids and transmembrane proteins, an immature spectrin--actin cytoskeleton, and residual organelles; over the $\sim\!24$--$48$~h that follow, vesiculation, pitting, and targeted proteolysis gradually remove this excess while cytoskeletal crosslinking matures~\cite{chasis1989membrane,mohandas2008red,Renoux2019Impact,chai2026pocked}. The multilobular, cup-shaped and near-discocytic stages therefore correspond to three well-defined mechanical states rather than to arbitrary morphological bins, and their simultaneous presence in the circulation means that even a modest reticulocyte fraction---$\sim\!0.5$--$2.5\%$ under normal conditions and higher during the erythropoietic response to sustained altitude~\cite{siebenmann2017regulation}---introduces a mechanically heterogeneous subpopulation whose flow behavior cannot be captured by a single cell model.

\begin{figure}[!tbp]
\begin{center}
\includegraphics[width=1.000\textwidth]{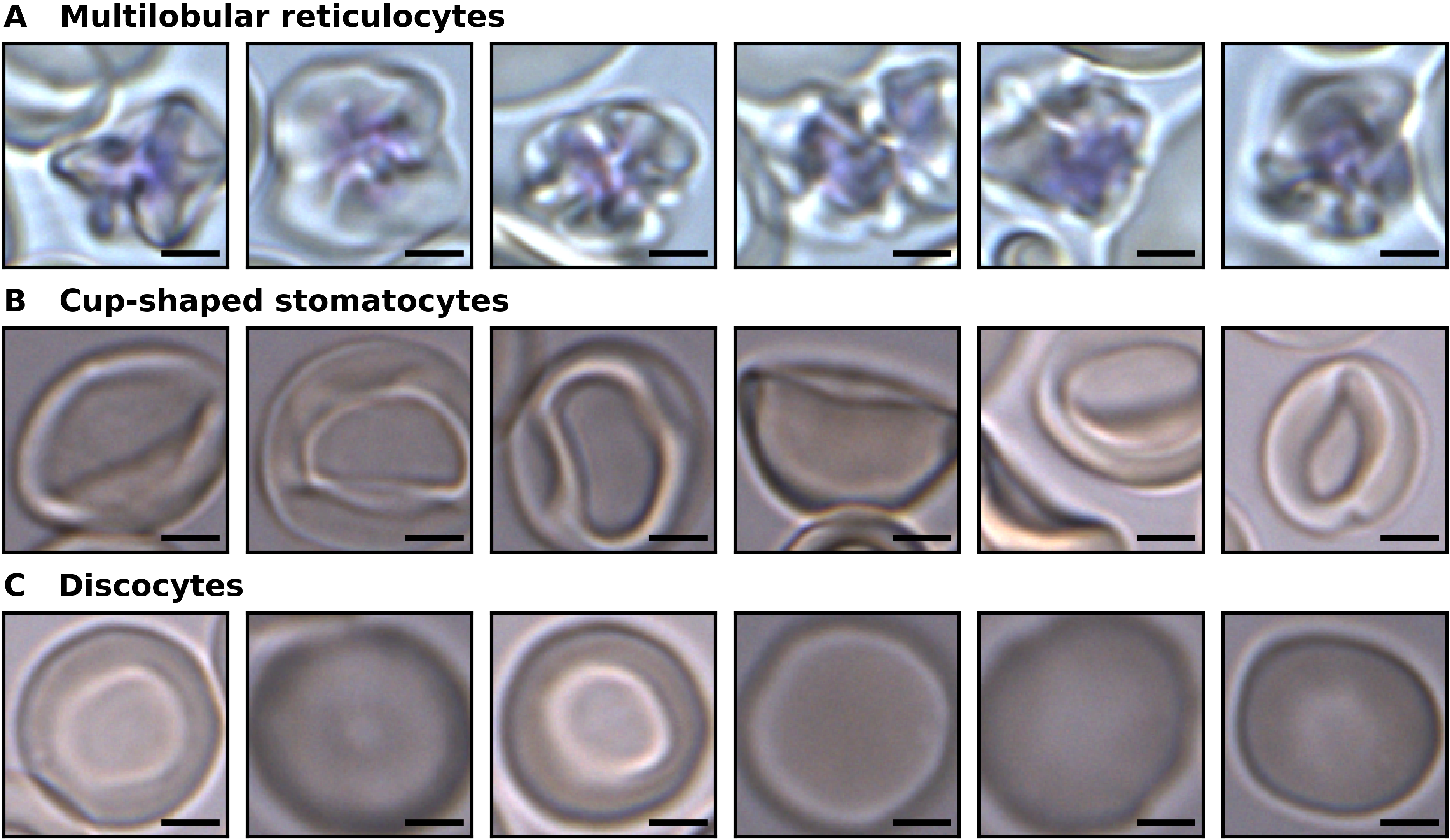}
\end{center}
\vspace{-0.15in}
\caption{\small{\bf Supravital NMB staining concentrates reticulum in multilobular cells; most cup-shaped cells are stomatocytes.}
All panels show cells selected as clear examples of their class; they are not the whole of any class. The full blind sample is tabulated in Table~\ref{tab:nmb_census}. Panel A cells are from the more strongly stained preparation, panels B and C from the blind sample.
(A) Multilobular reticulocytes, carrying a dense granulo-filamentous violet reticulum.
(B) Cup-shaped \emph{stomatocytes}~\textbf{---}~unilobular cup-shaped cells with no chromogen anywhere in the cell. Approximately $95\%$ of the cup-shaped cells in the blind sample are of this kind.
(C) Discocytes, with no chromogen.
Cup-shaped reticulocytes, the maturation stage modeled by R2, share the bright-field morphology of (B) and are not shown, because the sparse discrete granules of a late reticulocyte cannot be reliably distinguished by NMB from Howell--Jolly or Heinz bodies (\nameref{S1_Text}, Section 6). Scale bars $2~\si{\micro\metre}$.}
\label{fig:nmb_stain}
\end{figure}

Because these three classes are defined by cell shape alone, we asked directly whether they correspond to cells that still contain RNA. An additional donor sample was stained supravitally with New Methylene Blue and read under a two-pass design in which shape was classified from blue-channel images, where the chromogen is invisible, and the presence of reticulum was adjudicated separately from the color images (Fig~\ref{fig:nmb_stain}; Methods; \nameref{S1_Text}, Section 6). Unambiguous reticulum was concentrated in the multilobular class: of the eleven cells carrying it, ten were multilobular ($91\%$) and one was cup-shaped, and no discocyte carried any---$0$ of the $40$ scored here, and none among the $\sim\!5{,}200$ cells surveyed in a second, more strongly stained preparation of the same sample. Multilobular cells were positive far more often than discocytes (Table~\ref{tab:nmb_census})---$32.3\%$ versus $0\%$ for unambiguous reticulum (Fisher exact $p = 9.6\times10^{-5}$), and $67.7\%$ versus $2.5\%$ once equivocal calls are included ($p = 1.4\times10^{-9}$)---which supports the identification of R1 with genuinely immature, RNA-containing cells.

The cup-shaped class behaved differently. Only $1.5\%$ of cup-shaped cells carried unambiguous reticulum ($5.9\%$ including equivocal calls), a rate statistically indistinguishable from that of discocytes ($p = 1.0$ for unambiguous reticulum, $p = 0.65$ with equivocal calls included). In this sample, approximately $95\%$ of the cells that bright-field morphology places in the cup-shaped bin are therefore stomatocytes carrying no detectable RNA (Fig~\ref{fig:nmb_stain}B), and cup-shaped morphology by itself does not identify a reticulocyte. The more strongly stained preparation did contain cup-shaped cells with reticulum ($9$ of the $43$ reticulum-positive cells found in it), but it was surveyed in a targeted pass over stained cells, with shape recorded while the stain was in view, so it yields no class-conditional rate. Nor can supravital staining identify a late reticulocyte cell by cell: as maturation proceeds the reticulum is reduced to a few discrete granules~\cite{malleret2013significant,gaur2021reticulocyte}, and granules of this kind can be difficult to tell apart in a supravitally stained preparation from other red-cell inclusions such as Howell--Jolly or Heinz bodies~\cite{gaur2021reticulocyte} (\nameref{S1_Text}, Section 6, reports how such cells were handled and shows that no conclusion depends on it). The cup-shaped reticulocyte stage is accordingly taken from the maturation literature~\cite{mel1977reticulocyte,gronowicz1984maturation,chasis1989membrane,StevensHernandez2022Reticulocyte,Leduc2025Comprehensive} rather than from our staining, and R2 is a model of that stage, not of every cup-shaped cell encountered in a smear. The practical consequence for any morphology-only classification, including the one underlying Fig~\ref{fig:morphology}, is that the cup-shaped bin is diluted by stomatocytes and the reticulocyte fraction assigned to it is correspondingly overestimated. The mechanical descriptors of R2 are unaffected by this, because they were assigned from the reported mechanics of reticulocytes rather than from class counts. A complementary point applies to R3. That the near-discocytic class carried no unambiguous reticulum does not argue against a late reticulocyte stage but locates it: RNA clearance and membrane remodeling are not simultaneous, and the terminal phase of surface-area loss continues after the reticulum has become undetectable by supravital staining~\cite{chasis1989membrane,gronowicz1984maturation}. R3 is therefore expected to be stain-negative, and its descriptors were assigned from the reported mechanics of reticulocytes rather than from stain status. Neither late class, then, can be resolved by supravital staining: identifying cup-shaped or near-discocytic reticulocytes within a mixed population requires a molecular marker of immaturity, such as an RNA dye combined with the transferrin receptor CD71~\cite{Leduc2025Comprehensive}, and for R3 a mechanical assay.

\begin{figure}[!tbp]
\begin{center}
\includegraphics[width=1.00\textwidth]{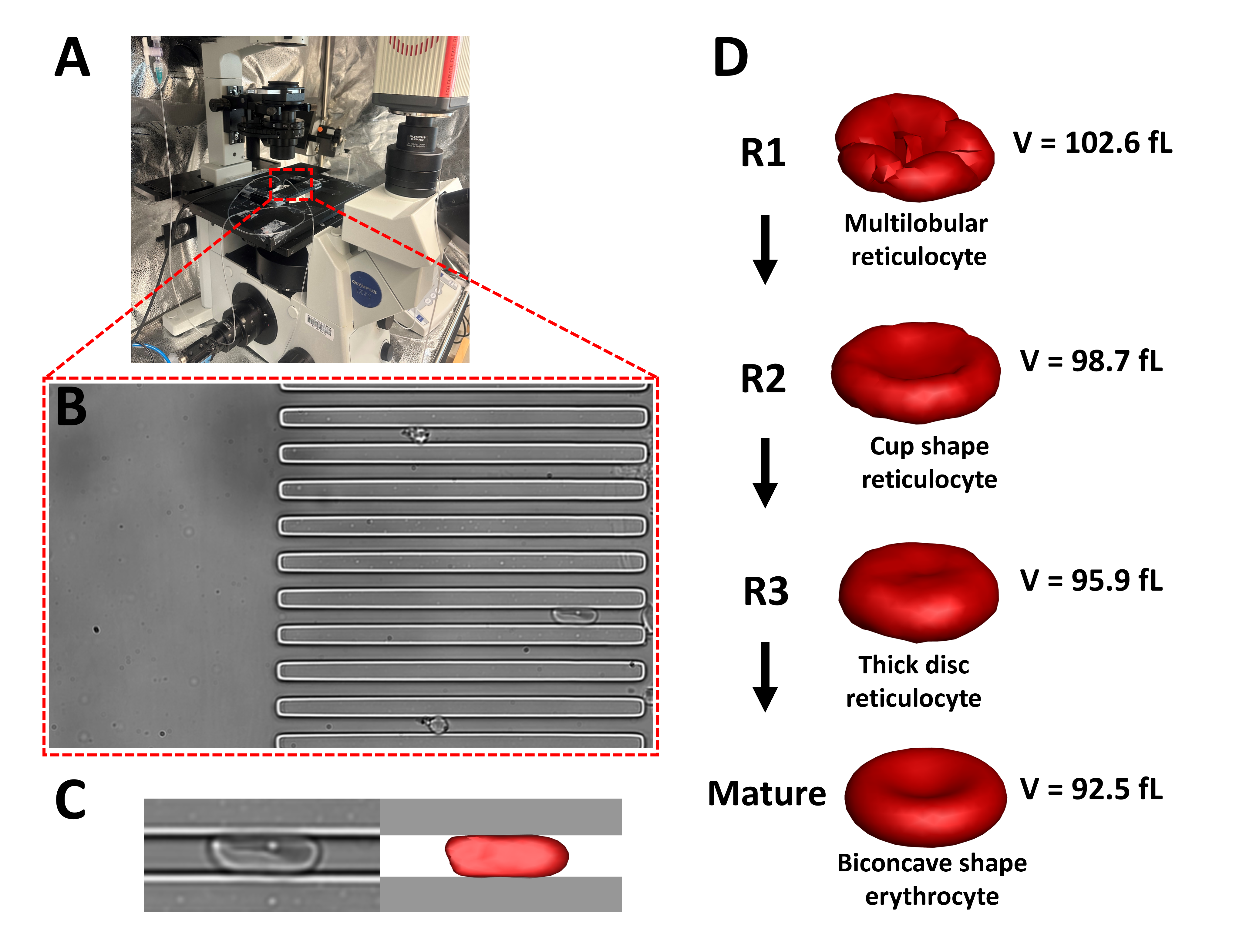}
\end{center}
\vspace{-0.1in}
\caption{\small{\bf Experimental microchannel platform and reticulocyte morphology spectrum.}
(A) Experimental microscopy setup used to image single-RBC transit through microfluidic channels.
(B) Bright-field view of the microfluidic microchannel array.
(C) Representative single-cell transit: experimental bright-field image (left) and the corresponding DPD simulation (right) of an RBC deforming through a channel.
(D) Reticulocyte maturation sequence---multilobular (R1), cup-shaped (R2), thick-disc (R3), and mature biconcave erythrocyte---with the reference volumes of the corresponding models ($102.6$, $98.7$, $95.9$, and $92.5$~fL; Table~\ref{tab:mechanical_properties}).}
\label{fig:platform}
\end{figure}

To represent this heterogeneity in our DPD simulations, we parameterized three reticulocyte models---R1, R2, R3---that map onto the multilobular, cup-shaped, and near-discocytic classes respectively (Fig~\ref{fig:platform}D; Methods; Table~\ref{tab:mechanical_properties}). The assigned descriptors span $\mu \in [6.3, 8.3]~\si{\micro\newton\per\metre}$ (factor $\sim\!1.3$--$1.8\times$ above CTR), $S/V \in [1.48, 1.56]~\si{\per\micro\metre}$ ($3$--$8\%$ above CTR), and $k_c \in [4.8, 7.2] \times 10^{-19}$~J ($2$--$3\times$ above CTR), all within experimentally reported ranges~\cite{cluitmans2012red,malleret2013significant,Renoux2019Impact,moreau2023physical}. These three descriptors are parametrically decoupled: R1 is the stiffest in shear and carries the largest excess area but shares the lowest bending modulus of the three with R2, whereas R3 is the softest in shear and the stiffest in bending. The shape-transition simulations of Fig~\ref{fig:shape_transition} use all three descriptors and are repeated with the shear modulus alone as a control (\nameref{S1_Text}, Section~8). The transport simulations that follow vary the shear modulus alone, holding the reference area, reference volume and bending modulus at their control-discocyte values, so there the $S/V$ and $k_c$ values assigned to the classes are varied only in the tests of \nameref{S1_Text}, Sections~5 and~7. In the limit $\mu \to \mu_0$, $S/V \to (S/V)_0$, and $k_c \to k_{c,0}$, the R1--R3 models collapse onto the CTR-RBC parameters used in prior work~\cite{fedosov2010multiscale,suresh2005connections,Li2018Mechanics}. This parameter set was fixed before any of the flow simulations below and held unchanged throughout, so those simulations show what the assigned moduli imply rather than being fitted to them.

The single-cell microchannel assay of Fig~\ref{fig:platform} provides the experimental setting for these models. Reticulocyte-rich samples were driven through $5\times5~\si{\micro\metre}$ PDMS microchannels (Fig~\ref{fig:platform}A,B) and individual cells were imaged as they deformed through the constriction; Fig~\ref{fig:platform}C sets one such cell beside a DPD cell in a constriction of the same $5~\si{\micro\metre}$ width ($30 \times 5 \times 2.7~\si{\micro\metre\cubed}$), the geometry of the microchannel simulations below. Fig~\ref{fig:platform}D places the models on the maturation sequence with the reference volumes of Table~\ref{tab:mechanical_properties}.

\begin{table}[!htbp]
\centering
\caption{{\bf Reticulum by morphological class in the blind read ($n = 139$).}
Morphology was scored from blue-channel images, in which the chromogen is invisible
(granule optical-density excess: green $+0.082$, red $+0.041$, blue $-0.006$), so shape was
classified blind to staining; staining was then adjudicated separately, in color (Methods).
$p$-values are Fisher exact tests against the discocyte class. Stain calls are reported as made in the blind read; the handling of
discrete, sharply bounded bodies, which NMB also stains in Howell--Jolly and Heinz inclusions, and a sensitivity analysis
are given in \nameref{S1_Text}, Section 6.}
\label{tab:nmb_census}
\resizebox{\textwidth}{!}{%
\begin{tabular}{lccccc}
\hline
\textbf{Class (scored blind)} & \textbf{$n$} & \textbf{Clear reticulum} & \textbf{Clear $+$ equivocal} & \multicolumn{2}{c}{\textbf{$p$ vs discocyte}} \\
 & & & & clear & clear$+$equiv. \\
\hline
Multilobular & 31 & 10 ($32.3\%$) & 21 ($67.7\%$) & $9.6\times10^{-5}$ & $1.4\times10^{-9}$ \\
Cup-shaped   & 68 & 1 ($1.5\%$)   & 4 ($5.9\%$)   & $1.0$ & $0.65$ \\
Discocyte    & 40 & 0 ($0\%$)     & 1 ($2.5\%$)   & --- & --- \\
\hline
\end{tabular}}
\end{table}

\subsection*{Flow-induced shape response of the reticulocyte models}

We next examined how the CTR and R1--R3 models deform in unconstricted pressure-driven flow, a loading mode distinct from the constrictions below. Donor-matched experimental shape-transition data for the individual reticulocyte subtypes are not available, so this analysis characterizes the models rather than validating them against experiment.

\begin{figure}[!tbp]
\begin{center}
\includegraphics[width=0.98\textwidth]{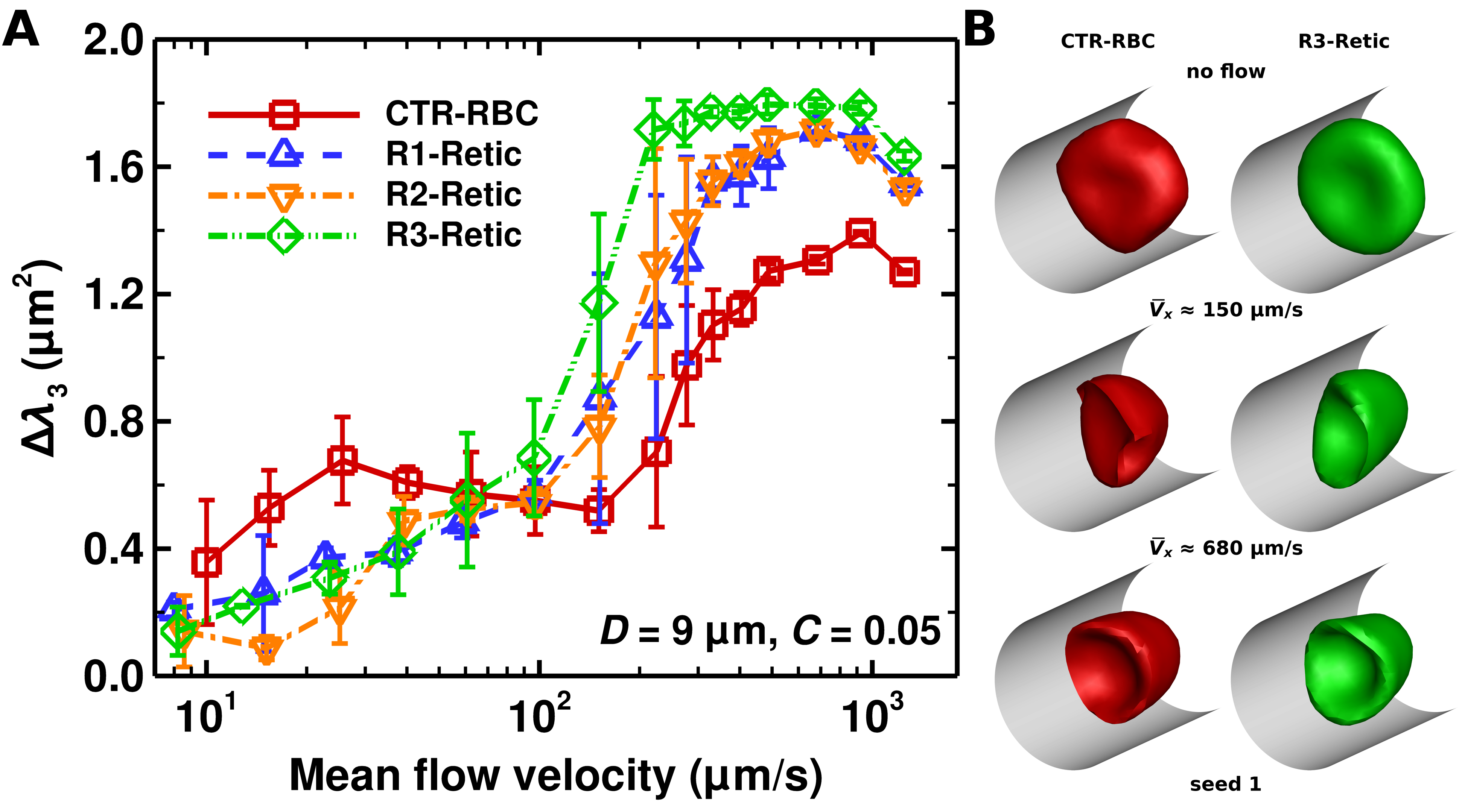}
\end{center}
\vspace{-0.1in}
\caption{\small{\bf Flow-induced shape transition quantified by the gyration tensor.}
(A) Shift of the smallest eigenvalue $\Delta\lambda_3$ of the gyration tensor from its zero-flow value as a function of mean flow velocity $\bar V_x$ for the CTR-RBC and the three reticulocyte models (R1--R3), each a single cell with the shear modulus, $S/V$ and bending modulus of Table~\ref{tab:mechanical_properties} in a periodic cylindrical channel of diameter $D = 9~\si{\micro\metre}$ at dilute volume fraction $C = 0.05$; mean $\pm$ SD of six realizations at $\bar V_x \gtrsim 60~\si{\micro\metre}/\mathrm{s}$ and of three at lower velocities and for the control at the highest velocity, each averaged over the second half of its run. The eigenvalue $\lambda_3$ measures the effective cell thickness along the vorticity direction and increases as the cell transitions from a biconcave discocyte to an axisymmetric parachute shape under flow. The reticulocyte models reach the parachute shape at lower velocities than the control and saturate higher, because they start thinner (Results); with the shear modulus alone varied, the four curves coincide (\nameref{S1_Text}, Section~8). (B) The control cell and R3 (one realization each; colors as in A) at rest, at $\bar V_x \approx 150~\si{\micro\metre}/\mathrm{s}$, where R3 is halfway through its transition and the control is not, and at $\bar V_x \approx 680~\si{\micro\metre}/\mathrm{s}$. Each cell is seen obliquely from upstream inside the far half of the channel wall (gray), with the flow directed away from the viewer.}
\label{fig:shape_transition}
\end{figure}

Specifically, we extracted a rotation-invariant measure of deformation under pressure-driven flow from the gyration tensor of each membrane configuration. For an RBC of $N_v$ membrane vertices at positions $\mathbf{r}_i = (x_i, y_i, z_i)$ with center-of-mass $\mathbf{r}_C = N_v^{-1} \sum_i \mathbf{r}_i$,
\begin{equation}
G_{\alpha\beta} = \frac{1}{N_v} \sum_{i=1}^{N_v} (r_i^\alpha - r_C^\alpha)(r_i^\beta - r_C^\beta),
\quad \alpha,\beta \in \{x,y,z\}.
\label{eq:gyration}
\end{equation}
Diagonalization of $\mathbf{G}$ yields three non-negative eigenvalues $\lambda_1 \geq \lambda_2 \geq \lambda_3$, whose square roots are the principal dimensions of the cell. The smallest eigenvalue $\lambda_3$ is a robust proxy for the effective cell thickness along the vorticity direction and is highly sensitive to the discocyte-to-parachute transition. We report the shifted quantity $\Delta\lambda_3(\bar V_x) = \lambda_3(\bar V_x) - \lambda_3^{\mathrm{eq}}$, with $\lambda_3^{\mathrm{eq}}$ extracted from a zero-flow reference, isolating the flow-induced deformation from equilibrium shape differences. Fig~\ref{fig:shape_transition}A shows $\Delta\lambda_3$ versus the mean flow velocity $\bar V_x$ for CTR-RBC and R1--R3, each simulated as a single cell in a $D = 9~\si{\micro\metre}$ cylindrical channel at dilute volume fraction $C = 0.05$ with the shear modulus, $S/V$ and bending modulus of Table~\ref{tab:mechanical_properties} (Methods). Every model ends in the parachute shape (Fig~\ref{fig:shape_transition}B), one of the shapes that simulations of red cells in cylindrical microchannels predict over a range of confinements and flow rates~\cite{fedosov2014deformation}, but the reticulocyte models reach it at lower flow velocities. Below $\bar V_x \approx 50~\si{\micro\metre}/\mathrm{s}$ they deform less than the control; the four models meet at $\Delta\lambda_3 \approx 0.5$--$0.6~\si{\micro\metre\squared}$ at $\bar V_x \approx 60$--$100~\si{\micro\metre}/\mathrm{s}$, where one of the six R3 realizations at each driving has already begun its transition, and the reticulocyte models leave this level first: at $\bar V_x \approx 150~\si{\micro\metre}/\mathrm{s}$ all six control realizations are still on it ($\Delta\lambda_3 = 0.52 \pm 0.07~\si{\micro\metre\squared}$), whereas R3 is halfway through its transition ($1.17 \pm 0.28~\si{\micro\metre\squared}$). The velocity at which $\Delta\lambda_3$ passes midway between that level and its saturated value, $\bar V_x^{c}$, is $267 \pm 38~\si{\micro\metre}/\mathrm{s}$ for the control and $154 \pm 21$, $207 \pm 43$ and $202 \pm 63~\si{\micro\metre}/\mathrm{s}$ for R3, R2 and R1 (mean $\pm$ SD of six realizations), $0.58$, $0.77$ and $0.76$ times the control value (Welch's $t$-test, $p = 0.0002$, $0.03$ and $0.06$). The reticulocyte models also saturate higher, at $1.6$--$1.8$ against $1.3~\si{\micro\metre\squared}$, because they start thinner: with less volume per unit membrane area, their zero-flow $\lambda_3$ is $0.73$--$0.80~\si{\micro\metre\squared}$, against $1.19~\si{\micro\metre\squared}$ for the control, whereas the parachute has nearly the same $\lambda_3$ ($2.41$--$2.58~\si{\micro\metre\squared}$) in all four. The higher saturation raises the level at which their $\bar V_x^{c}$ is read, so it works against the earlier transition rather than producing it. Nor does the shear modulus produce it. With every model at the reference volume and bending modulus of the control, so that only the shear modulus differs, the four models follow a common curve (\nameref{S1_Text}, Section~8): $\bar V_x^{c}$ is $264 \pm 42$, $272 \pm 48$ and $287 \pm 63~\si{\micro\metre}/\mathrm{s}$ for R3, R2 and R1 ($p \geq 0.5$ against the control), and R1, whose shear modulus is $1.75$ times that of the control, reaches the parachute shape at a velocity $7\%$ higher in the mean, within the realization-to-realization spread. The earlier transition of the reticulocyte models therefore comes from their lower volume per unit membrane area and larger bending modulus, whose contributions these runs do not separate, and not from their shear modulus.

\subsection*{Simulations of single-cell transit dynamics through confined geometries}

\begin{figure}[!tbp]
\begin{center}
\includegraphics[width=0.98\textwidth]{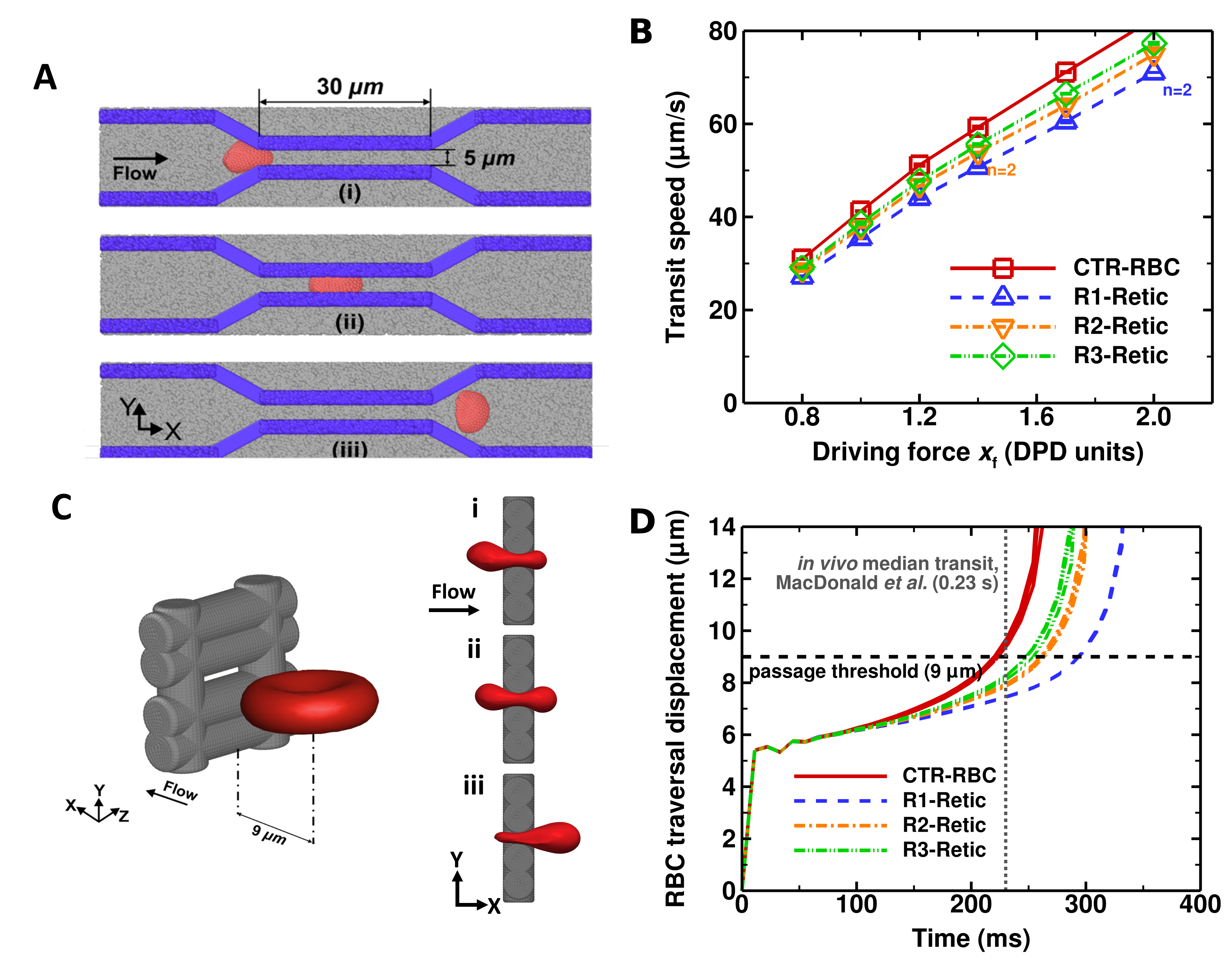}
\end{center}
\vspace{-0.1in}
\caption{\small{\bf Differential sensitivity of reticulocyte transit in microchannel and inter-endothelial-slit (IES) geometries.}
(A) Schematic of the DPD simulation domain for the microchannel (width $5~\si{\micro\metre}$, length $30~\si{\micro\metre}$), showing a single RBC entering, traversing, and exiting the constriction under pressure-driven flow.
(B) Speed of the cell's center of mass through the constriction (mean over the stretch from $5$ to $20~\si{\micro\metre}$ past its entrance) as a function of the driving body force $x_f$ (DPD units; Methods) for CTR-RBC and R1--R3; mean $\pm$ SD over three realizations, and over four for R1 at $x_f = 0.8$; two runs were lost to a numerical membrane rupture inside the constriction (R1 at $x_f = 2.0$, R2 at $x_f = 1.4$), and the number of realizations behind any point averaged over fewer than three is printed beside it. The reticulocyte models are slower in the order of their shear modulus ($\mu$: R3 $<$ R2 $<$ R1), by at most $\sim\!15\%$.
(C) Inter-endothelial-slit (IES) configuration built from parallel cylindrical endothelial elements (gray). Left: the three-dimensional arrangement, with an RBC (red) positioned upstream of the slit; the dash-dotted markers indicate the $9~\si{\micro\metre}$ between the cell's initial center of mass and the plane of the slit, through the axes of the cylinders that form it; this is the passage threshold of panel (D). Right: three sequential snapshots of a single CTR-RBC squeezing through the same slit, viewed in the $x$--$y$ plane with flow left to right---(i) the cell already engaged in the slit entrance with a thin lobe protruding downstream, (ii) the cell midway through, adopting a dumbbell shape whose waist is pinched by the aperture, and (iii) the cell emerging downstream while its trailing lobe is still within the slit.
(D) Displacement of the cell's center of mass along the flow direction versus time after release, for three realizations of each cell type (one R2 realization, lost to a numerical membrane rupture after its center of mass had passed the slit plane, is replaced by a fourth). The dashed line at $9~\si{\micro\metre}$ marks the center of mass reaching the plane of the slit (panel C); the dotted line marks the median in vivo transit time through rat splenic slits, $0.23$~s (MacDonald \textit{et al.}~\cite{macdonald1987kinetics}; range $0.02$--$60.5$~s). The advance of $\sim\!5~\si{\micro\metre}$ within the first $\sim\!10$~ms, and the brief recoil after it, are the start-up of the driven solvent before pressure builds up against the slit; the steep rise after passage is the cell being drawn into the rarefied solvent downstream (\nameref{S1_Text}, Section 5). The slit is driven at a constant trans-slit pressure gradient of $6$~Pa/\si{\micro\metre}, as in Liu \textit{et al.}~\cite{LIU20214heated}, and times use the conversion of Methods, which is fixed by the control transit through this slit. R3, R2 and R1 reach the slit plane $13$, $18$ and $33\%$ later than the control, whereas in the microchannel of panel (B) R1 takes at most $\sim\!18\%$ longer than the control.
The two geometries were each driven at the loading appropriate to them rather than at a common gradient, so these spreads are compared as within-geometry contrasts and not as an equal-load measurement.}
\label{fig:microchannel}
\end{figure}

We next examined how reticulocyte shape and deformability govern transit through two physiologically distinct confined geometries: a narrow microchannel ($5~\si{\micro\metre}$ wide, $2.7~\si{\micro\metre}$ deep and $30~\si{\micro\metre}$ long) that emulates the tightest capillary-scale constrictions encountered in the microcirculation, and a splenic inter-endothelial-slit (IES)-like gap ($1.5~\si{\micro\metre} \times 5.5~\si{\micro\metre}$ cross-section, $2.5~\si{\micro\metre}$ deep) that mimics the size-selective barrier separating splenic cords from sinusoids (Fig~\ref{fig:microchannel}A,C). Each configuration was driven at the loading appropriate to it: the microchannel by a uniform body force swept over $x_f = 0.8$--$2.0$ in DPD units (Methods), and the slit at the constant trans-slit gradient of $6$~Pa/\si{\micro\metre} that Liu \textit{et al.}~\cite{LIU20214heated} applied to this slit geometry. Within each geometry the four cell types were driven identically, so the subtype spread that each geometry resolves is internally controlled; the two geometries were not driven at the same loading, and are therefore compared through those within-geometry spreads rather than through absolute velocities. In the microchannel (Fig~\ref{fig:microchannel}B), transit depended only weakly on the cell's mechanics. The control discocyte crossed the constriction at $31$--$83~\si{\micro\metre}/\mathrm{s}$ over the sweep, and the reticulocyte models were slower in the order of their shear modulus, R1 $<$ R2 $<$ R3 $\lesssim$ CTR, but by small margins: R1 took $14$--$18\%$ longer than the control to cross the constriction and R3 at most $7\%$ longer, the separation being largest at $x_f = 1.7$.

In the narrower IES-like slit ($1.5~\si{\micro\metre}$ gap versus the $5~\si{\micro\metre}$ microchannel; Fig~\ref{fig:microchannel}C) the same models were separated nearly twice as strongly (Fig~\ref{fig:microchannel}D). The center of mass of the control discocyte, released $9~\si{\micro\metre}$ upstream of the plane of the slit---the plane through the axes of the cylinders that form it, where the gap is narrowest---reached that plane after $222 \pm 2$~ms and a plane $2.5~\si{\micro\metre}$ further downstream, beyond the cylinders, after $249 \pm 3$~ms; R3, R2 and R1 reached the slit plane $13$, $18$ and $33\%$ later ($250 \pm 4$, $262 \pm 2$ and $294 \pm 1$~ms; mean $\pm$ SD of three realizations) and the downstream plane $11$, $16$ and $29\%$ later. The delay follows the shear modulus, the only parameter that differs between these models, and is nearly twice the largest delay that the microchannel produced at any driving. It is incurred at the aperture: the displacement curves of the four cell types coincide over the first $\sim\!6~\si{\micro\metre}$ of the approach and part only as the cell engages the slit. All four cells nevertheless cleared the slit at this loading, $\sim\!250$--$320$~ms after release. These times lie near the median of the in vivo transit times through rat splenic slits measured by MacDonald \textit{et al.}~\cite{macdonald1987kinetics} ($0.23$~s; range $0.02$--$60.5$~s), but because the time scale of this work is itself fixed by the control transit through this slit (Methods), the agreement places the reticulocyte delays within the in vivo distribution rather than testing the loading independently. A comparable splenic-slit setting applied to Gaucher-disease red cells, whose rigidity is far greater, extends the passage time beyond $1200$~ms~\cite{chai2026gaucher}, placing the reticulocyte subtypes at the compliant end of a common scale. Because the transport runs hold every model at the control-discocyte surface-to-volume ratio, we repeated the slit runs with the reference volume of each model lowered, at the control membrane area, to give it the $S/V$ of Table~\ref{tab:mechanical_properties}; the delays at the slit plane shrank to $9$, $15$ and $26\%$ but kept their order (\nameref{S1_Text}, Section 5). Whether a cell passes an inter-endothelial slit at all is set chiefly by its surface area and volume~\cite{pivkin2016biomechanics,dao2021erythrocyte}; every model here carries enough of both to pass, and what its shear modulus sets is how long the passage takes. That the narrow slit resolves this more sharply than the microchannel is expected, since it forces a far larger membrane deformation, and it is consistent with the sensitivity of the spleen to reduced red-cell deformability~\cite{safeukui2018sensing,qiang2023microfluidic}.

Taken together, the two geometries order the reticulocyte models in the same way, by shear modulus, but with different sensitivity: in the $5~\si{\micro\metre}$ microchannel R1 takes at most $\sim\!18\%$ longer than the control to cross the constriction, whereas the $1.5~\si{\micro\metre}$ slit delays it by $\sim\!30\%$. Capillary-scale transit therefore carries only a small penalty for an immature reticulocyte, and the splenic slit is where its stiffness matters. At the trans-slit loading used here that penalty is a delay rather than retention---every model cleared the slit---so whether the spleen retains a given population is decided by its critical driving $\Delta P_c$, quantified below; this is consistent with splenic release during acute hypoxia recruiting the full R1--R3 spectrum without substantial preferential retention, a point that re-emerges in the acute-exposure box of Fig~\ref{fig:conceptual}C.

\subsection*{Simulations of the impact of cell--cell interaction on the RBC transit dynamics through a slit}

\begin{figure}[!tbp]
\begin{center}
\includegraphics[width=1.00\textwidth]{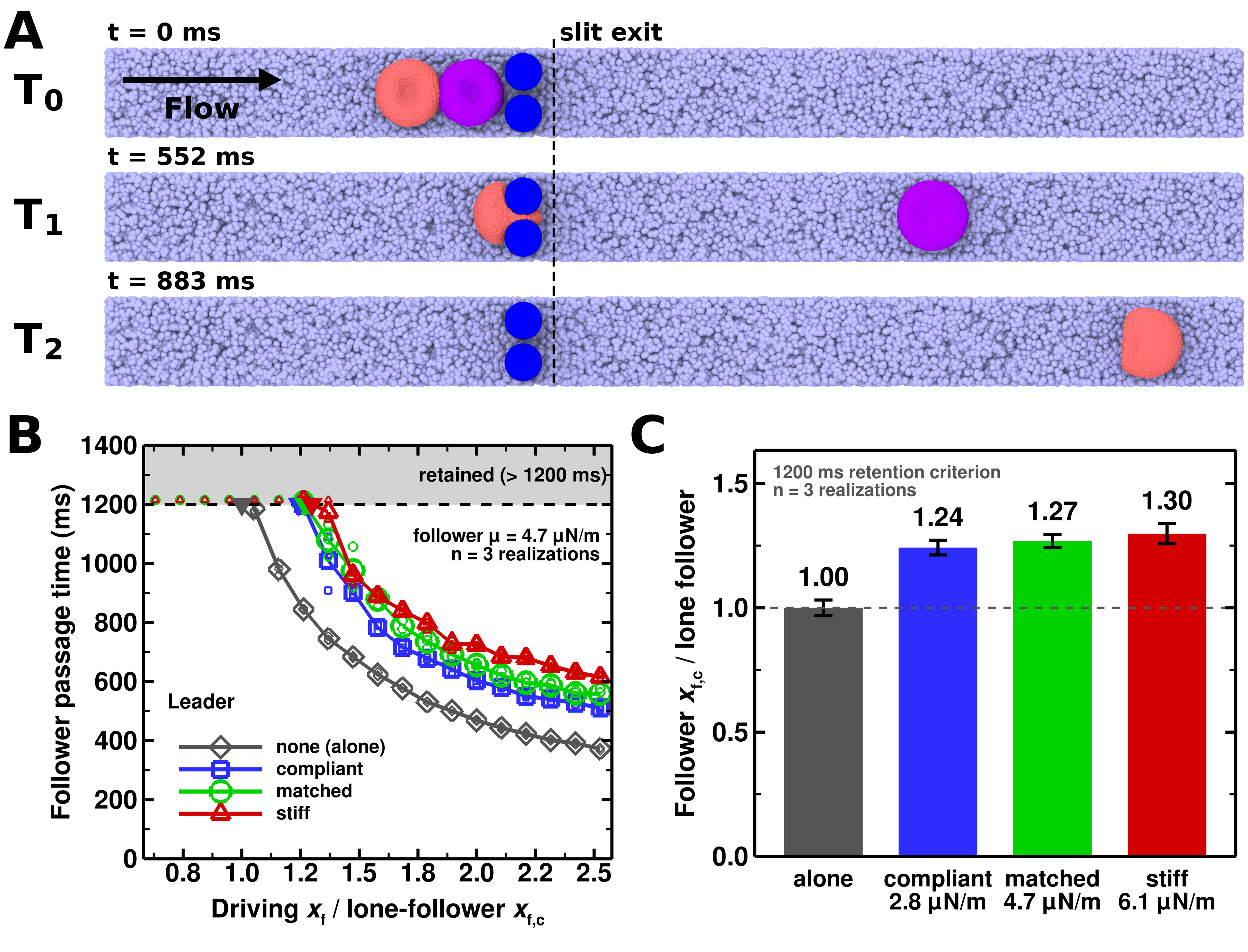}
\end{center}
\vspace{-0.1in}
\caption{\small{\bf A leading cell delays, and never rescues, the cell behind it.}
(A) DPD snapshots (top view; fluid not shown) of a two-cell train---a leader (magenta) followed by a trailing cell (red)---driven left-to-right (flow arrow) in single file through the physiologically scaled splenic slit between two posts (blue; net height comparable to the cell thickness), here behind the matched leader at a driving $1.7\times$ the lone follower's critical driving. At $T_0$ both cells sit upstream of the slit; at $T_1$ the leader has cleared the slit and the trailing cell is crossing it; by $T_2$ the trailing cell has deformed through the pore and advanced downstream. The time since release is given above each row.
(B) Passage time of the trailing cell (its center of mass reaching the slit exit) against the driving $x_f$, in units of the lone follower's critical driving $x_{f,c}$, alone and behind three leaders spanning a shear-modulus ladder---$\mu = 2.8$, $4.7$ and $6.1~\si{\micro\newton\per\metre}$, i.e.\ $0.6$, $1.0$ and $1.3\times$ the follower's own $\mu = 4.73~\si{\micro\newton\per\metre}$ (the stiffest leader additionally carries $5\%$ more reference area; all other membrane parameters are those of the control discocyte throughout). Lines join the averages over $n=3$ independent realizations, from which the thresholds are computed; small symbols are single realizations, and a run in which the cell was retained throughout sits at the run length, in the shaded band above the dashed $1200$~ms retention criterion. The filled markers on the criterion give each critical driving.
(C) Critical driving of the trailing cell, alone and behind each leader, relative to the lone follower (bar\,$\pm$\,SD of the per-realization thresholds, $n=3$): every leader raises it, the stiffest the most ($1.30\times$, against $1.24\times$ behind the most compliant leader). Because only the leader's shear modulus is varied, the comparison isolates the effect of a neighbor's deformability and is not specific to any one cell class. A leader never lowers the follower below its own single-cell threshold (no rescue); its compliance sets only how long the follower waits.}
\label{fig:pairwise}
\end{figure}

To ask whether a compliant cell can actively assist a stiffer neighbor, we performed pairwise DPD simulations in which two cells were driven in single file through a narrow downstream constriction of width $1.5~\si{\micro\metre}$---narrower than the RBC thickness and comparable to the scale of splenic inter-endothelial slits---and swept the driving body force across the clogging transition (Fig~\ref{fig:pairwise}). Cell transit in this regime is governed by viscous and membrane stresses, and a short-range repulsive contact term kept the membranes from overlapping one another and the posts that form the slit (Methods). Passage was scored as for the single-cell thresholds of Fig~\ref{fig:dpc_ladder}: a cell is retained if its center of mass has not reached the slit exit within the $1200$~ms retention criterion, and the critical driving is the driving at which the passage time, averaged over three independent realizations, crosses that criterion. In the single-cell reference, a lone cell carrying the control discocyte's shear modulus ($\mu = 4.73~\si{\micro\newton\per\metre}$, with its reference area, reference volume and bending modulus) exhibited a finite critical driving: it remained lodged at the constriction entrance at low drive but, once the driving force exceeded a critical value, deformed and broke through, clearing the pore and advancing far downstream~\cite{quinn2011combined,perazzo2022effect}. The control rung of the single-cell ladder of Fig~\ref{fig:dpc_ladder} is this same configuration measured with the particle contact term alone, and the two thresholds agree to within their uncertainties (Methods).

When a more compliant cell ($\mu = 2.8~\si{\micro\newton\per\metre}$, $0.6\times$ the follower's modulus) was placed ahead of it and the pair driven together, we first asked whether the leader lets the follower pass \emph{below its own} isolated threshold. It does not. Across the $19$ driving forces and three independent realizations swept for each leader, no follower cleared the constriction within the criterion at a driving at which all three lone followers were retained, and at no driving was its realization-averaged passage time behind a leader shorter than alone. On the contrary, in a single-file slit only wide enough to admit one cell, a leading cell \emph{raises} the follower's critical driving above its isolated value: relative to the threshold the same cell requires when driven alone through the identical slit under the identical protocol, the three leaders lift it by $24$--$30\%$ (Fig~\ref{fig:pairwise}C). The trajectories show why. The follower reaches the slit entrance only as the leader clears the exit---within $25$~ms of it in every one of the $111$ runs in which both cells passed---and then crosses in about the time it needs alone ($1.02\times$ its lone transit time at the same driving, on average; Fig~\ref{fig:pairwise}A). The pair passes as a queue: the follower's passage time is the leader's passage time plus the follower's own transit, to within $2\%$ on average. Behind the compliant, matched and stiff leaders the crossing itself took $1.06$, $1.01$ and $1.00\times$ the lone follower's transit time, shorter the longer the follower had waited (Pearson $r = -0.27$ between this ratio and the delay of the follower's arrival relative to the lone follower, over $111$ runs); these runs do not separate a wake effect from the driving head that is still rising while the cells cross (Discussion). A wake-mediated ``unjamming'' picture, in which a compliant leader would reduce a stiff neighbor's threshold by an order of magnitude, is therefore not supported---a cell passes most easily alone.

The leader's stiffness sets how long the follower waits, and this is where deformability matters. In the physiologically scaled splenic slit---a net channel height comparable to the cell thickness, so that the cell is confined against both walls and centered on the sub-cellular pore---we measured the critical driving at which the trailing cell cleared the constriction behind leaders spanning a shear-modulus ladder---$\mu = 2.8$, $4.7$ and $6.1~\si{\micro\newton\per\metre}$, that is $0.6$, $1.0$ and $1.3\times$ the follower's own modulus, the stiffest leader additionally carrying $5\%$ more reference area. A stiffer leader takes longer to cross the slit at every driving at which all runs passed (at the lowest of them, $1.47\times$ the lone follower's threshold, the three leaders cleared the exit after $421$, $535$ and $571$~ms on average), and the follower's threshold rose with leader stiffness accordingly, to $1.24$, $1.27$ and $1.30\times$ the lone follower's threshold, so that behind the stiffest leader the follower required a driving $\sim\!5\%$ higher than behind the most compliant one. At each of the $11$ drivings at which every run passed, the follower's passage times behind the three leaders spanned a narrower range than the leaders' own ($49$--$97\%$ of it), partly through the shorter crossing after a longer wait. Because this penalty is a waiting time rather than a mechanical barrier, its size in units of driving depends on how long the follower is given to pass: scored within a shorter observation window of $663$~ms, the same runs place the three thresholds $22$--$51\%$ above the lone follower's. Thus, although a leading cell never rescues a follower below its single-cell threshold, a \emph{compliant} leader, which vacates the aperture sooner, is a less obstructive neighbor than a stiff one in crowded single-file flow. These pairwise runs vary the leader's shear modulus alone and are therefore not tied to a particular cell class; they isolate how a neighbor's deformability, rather than its identity, sets the delay it imposes.

The physiological reading is correspondingly nuanced. Whether a given cell passes at all is set by its \emph{single-cell} critical pressure gradient $\Delta P_c$; cell--cell interaction does not override this. But in the queued, single-file passage of the splenic cords at elevated hematocrit, a cell's passage time includes the time the cell \emph{ahead} takes to clear the slit, so the compliance of that cell matters: replacing a stiff leader with a compliant one shortens the wait it imposes on the cells behind it and lowers their threshold by $\sim\!5\%$, easing collective throughput without altering any single cell's intrinsic $\Delta P_c$~\cite{recktenwald2024morphology,qiang2023microfluidic,chai2025silico}. The benefit a population confers at altitude is thus a combination of its own single-cell $\Delta P_c$ and this leader-compliance effect on the cells that follow, rather than a wake-mediated rescue of individually clogged stiff cells.

Taken together, the single-cell thresholds and the pairwise results define the local end of a geometry--mechanics map: whether a cell is retained is set by its own $\Delta P_c$, which a leading neighbor can raise but, in the configurations simulated, never lower. In the physiological setting of the splenic cord, this is consistent with a single spleen that (i) retains a subset of mechanically compromised cells at its inter-endothelial slits, (ii) slows their near neighbors, which may promote remodeling through vesiculation/pitting~\cite{pivkin2016biomechanics,moreau2023physical,li2021How}, and (iii) passes the rest of the RBC population, whose thresholds lie below the prevailing gradient---the spleen operating as a graded, rather than binary, filter.

\subsection*{Simulations of the critical driving for slit passage across RBC populations}

\begin{figure}[!tbp]
\begin{center}
\includegraphics[width=\textwidth]{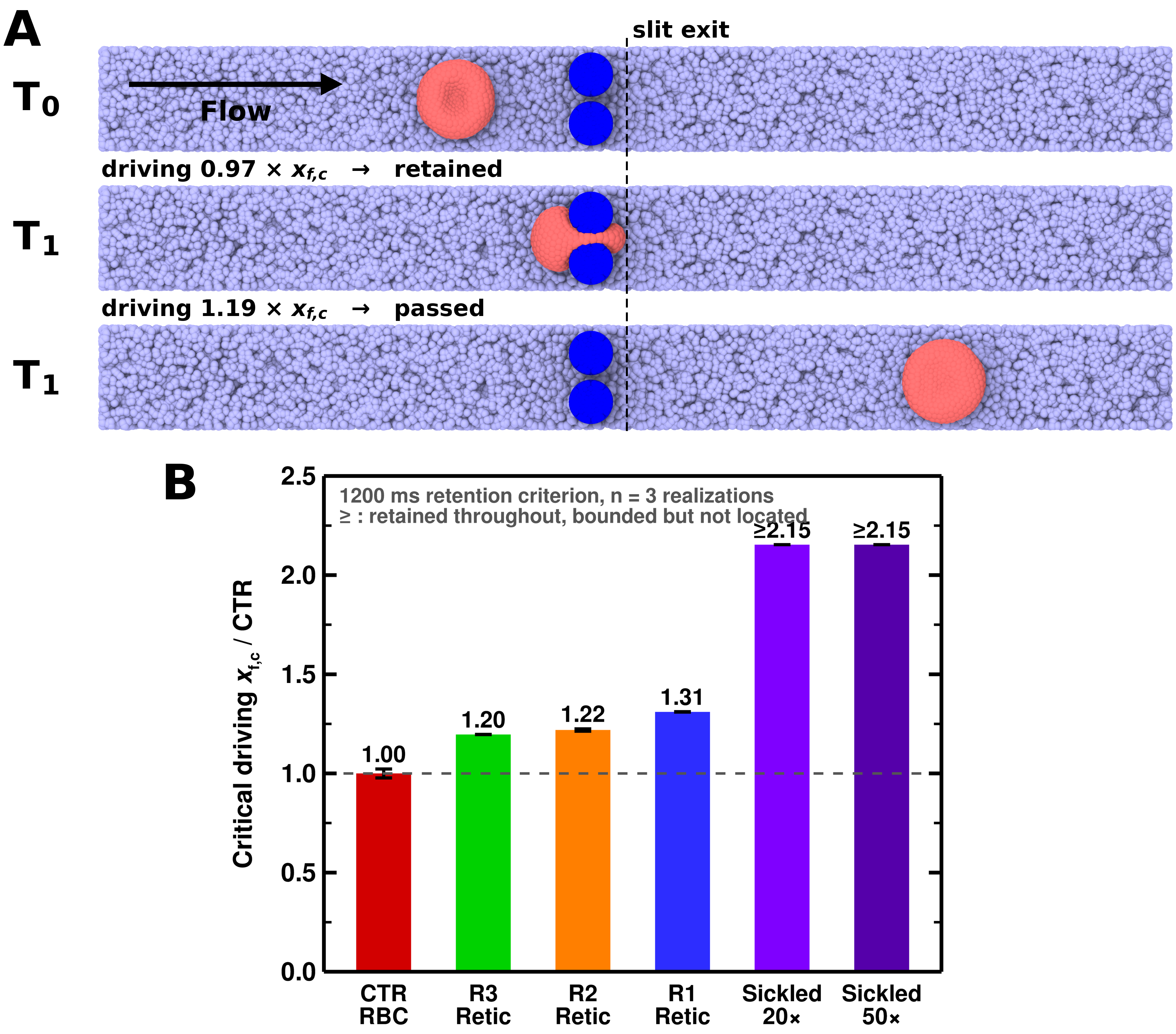}
\end{center}
\vspace{-0.1in}
\caption{\small{\bf Single-cell critical driving for slit passage across RBC populations.}
(A) DPD snapshots of the assay in top view, flow from left to right (arrow): a single cell (red) is driven through a slit whose two posts (blue) span the channel height and leave a $1.5~\si{\micro\metre}$ pore between them (a second, identical pore straddles the transverse periodic boundary at the strip edges; Methods); the background is the bottom channel wall, and the solvent and the top wall are not drawn. $T_0$: a control discocyte at release; $T_1$: $1100$~ms after release, in runs driven at $0.97\times$ (middle) and $1.19\times$ (bottom) the control's critical driving. Below the threshold the cell is still in the pore at $T_1$ and has not reached the slit-exit plane (dashed) when the run ends at $1215$~ms, past the $1200$~ms retention criterion; above it the cell cleared that plane at $997$~ms.
(B) Critical driving $\Delta P_c$ required for slit passage, relative to the CTR discocyte, for CTR-RBC, reticulocytes at three maturation stages (R1--R3) and sickled cells at $20\times$ and $50\times$ the control shear modulus (moduli of Table~\ref{tab:mechanical_properties} for CTR and R1--R3; membrane area, volume and bending modulus held at the control values, so the ladder isolates the shear modulus). These thresholds were measured in the channel of (A), an elongated slit channel of the type used for the pairwise transport assay (Methods). A cell is scored as passing if it clears the slit-exit plane within the $1200$~ms retention criterion; the threshold is the driving at which the passage time crosses that criterion, interpolated between bracketing runs $0.05$ apart in driving, with three independent realizations at every point (bars, realization-averaged threshold; error bars, standard deviation of the per-realization thresholds; values printed above the bars). The axis is the dimensionless driving of the assay (Methods); the ratios are the quantities to be compared across populations. Sickled cells exhibit the highest threshold: both were still retained at $x_f = 1.00$ in every realization, so their thresholds exceed $2.15\times$ that of CTR discocytes and are drawn as bounds (marked $\geq$) rather than located values (Methods).}
\label{fig:dpc_ladder}
\end{figure}

We next quantified, for each RBC population, the critical driving at which a single cell overcomes confinement-induced trapping at a slit (Fig~\ref{fig:dpc_ladder}A). By scoring slit passage against the $1200$~ms retention criterion used throughout this work, we identified distinct critical driving thresholds $\Delta P_c$ for each cell type (Fig~\ref{fig:dpc_ladder}B; three independent realizations at every bracketing driving). CTR discocytes exhibited the lowest $\Delta P_c$, followed in order by R3, R2 and R1 reticulocytes, with sickled cells at the highest end. Because the pressure scale of a driven DPD constriction is geometry dependent, we quote these thresholds relative to the CTR discocyte threshold measured in the same geometry rather than as absolute gradients: R3 $1.20\times$, R2 $1.22\times$ and R1 $1.31\times$ the CTR value. Neither sickled cell was located: at $20\times$ and at $50\times$ the control shear modulus every realization was still retained at $x_f = 1.00$, so both thresholds exceed $2.15\times$ the CTR value. The reticulocyte subtypes therefore sit $20$--$31\%$ above the control baseline, while the sickled-cell threshold exceeds $2.15\times$ that of CTR and $1.64\times$ that of the stiffest reticulocyte. Two features of this ladder deserve comment. First, the two sickled rungs are bounded rather than measured. Both membranes were still retained at $x_f = 1.00$ in every realization, and in the single realization run at $x_f = 1.20$ as well, so the ladder places both thresholds above $2.15\times$ the CTR value without locating either. Their near-equality at $20\times$ and $50\times$ the control modulus reflects that shared bound and cannot be read as the threshold ceasing to rise with stiffness. Second, the moduli assigned to the sickled cell, $94.6$ and $236~\si{\micro\newton\per\metre}$, bracket the mean shear modulus measured for irreversibly sickled cells, $135~\si{\micro\newton\per\metre}$, about $20$ times that of non-sickled cells, and cells that carry hemoglobin polymer are stiffer still ($801~\si{\micro\newton\per\metre}$)~\cite{Williams2025Quantifying}. HbAS cells that have not sickled, by contrast, stiffen by only $\sim\!1.3$--$1.8\times$ even under simulated strenuous exercise~\cite{Xu2016Stiffening} and thus fall within the reticulocyte band of Fig~\ref{fig:dpc_ladder}B. The mechanical distinction that matters at the slit is therefore not the genotype but whether a given cell has sickled. Once a population's $\Delta P_c$ approaches or exceeds the prevailing splenic trans-slit pressure, the cord-to-sinusoid passage ceases to be a probabilistic filter and becomes a near-deterministic trap for that population---and of the populations simulated here, the sickled cell reaches that regime first---providing a mechanical basis for the clinical observation that rapid altitude ascent in sickle-cell-trait (SCT) individuals, in whom a subpopulation of cells sickles under profound hypoxia, can produce acute splenic infarction~\cite{goodman2014splenic}.

The distribution of single-cell $\Delta P_c$ values allows a quantitative reading of the altitude syndromes considered in this study. In SCT individuals under the hypoxia of altitude, the cells that have sickled carry the highest threshold of the populations simulated---more than $2.15\times$ that of CTR discocytes and $1.64\times$ that of the stiffest reticulocyte---so they are the first population the spleen is biased toward trapping; this is the mechanical signature of acute splenic syndrome. In chronic mountain sickness (CMS), by contrast, the circulating population is enriched in R1--R3 reticulocytes, whose thresholds sit only $20$--$31\%$ above the CTR baseline and thus far below the sickled-cell value; splenic transit is preserved even at elevated hematocrit, but the downstream cost is the elevated low-shear viscosity documented in Fig~\ref{fig:viscosity}. Because every population within an individual is presented to the same splenic driving pressure, this separation is governed by the spacing of the thresholds rather than by their absolute values: at any operating point lying between the stiffest reticulocyte threshold and the sickled-cell threshold, sickled cells are retained while controls, reticulocytes and unsickled HbAS cells pass, so the split between the two phenotypes does not require the in vivo trans-slit gradient to be known precisely. The threshold ladder thus pins down the specific cell-level parameter---$\Delta P_c$ relative to the splenic trans-slit pressure---that separates the two clinical phenotypes, and the pairwise analysis above shows that tandem cell--cell interaction does not override it.

\subsection*{Blood viscosity under variable shear rates across physiological and pathological conditions}

\begin{figure}[!tbp]
\begin{center}
\includegraphics[width=0.85\textwidth]{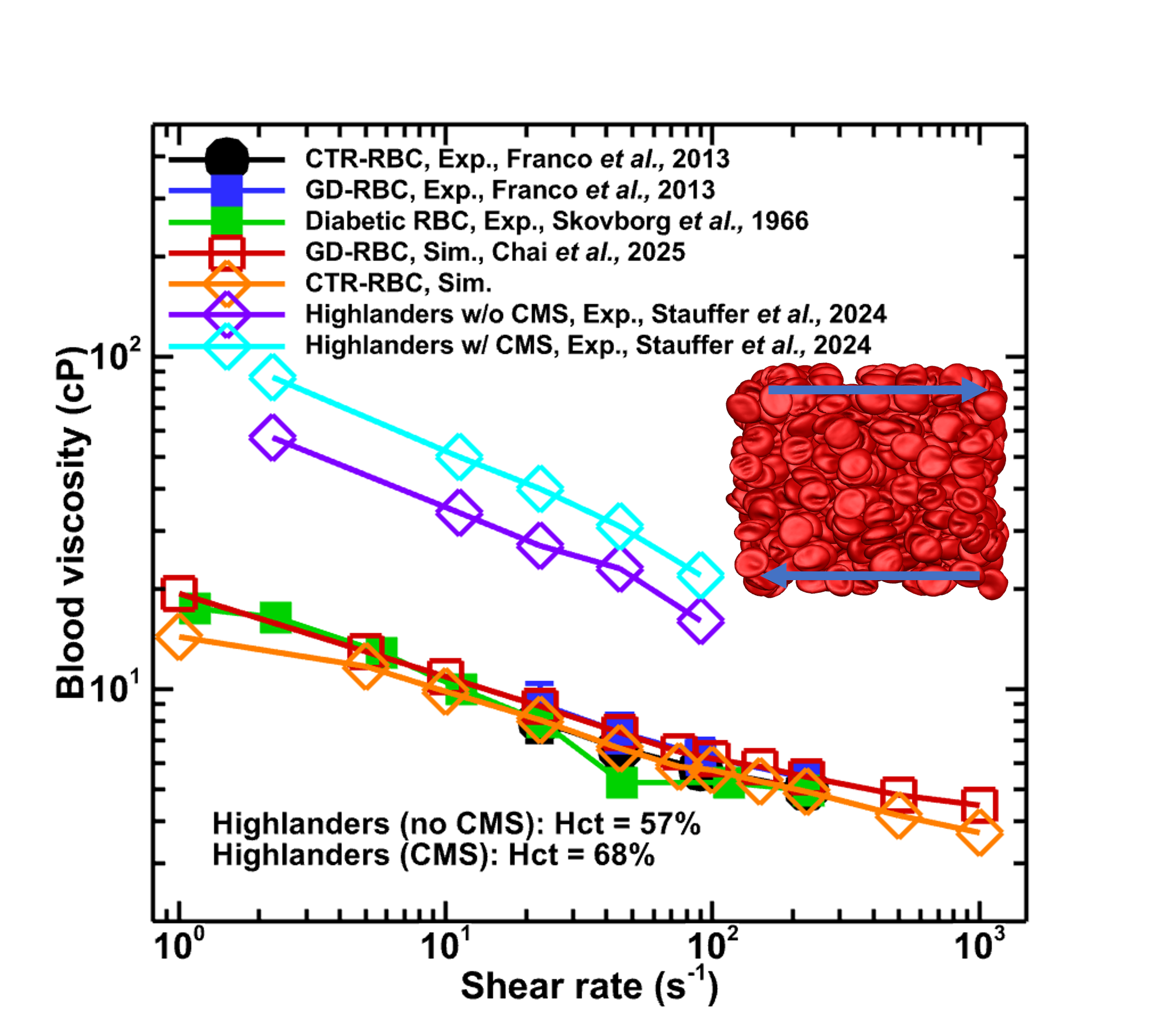}
\end{center}
\vspace{-0.1in}
\caption{\small{\bf Shear-rate-dependent blood viscosity across physiological and pathological conditions.}
Apparent blood viscosity $\eta(\dot\gamma)$ as a function of shear rate for CTR-RBC, Gaucher-disease RBCs (GD-RBC), diabetic RBCs, and high-altitude cohorts, compiled from published experiments and compared with DPD simulations. Experimental CTR- and GD-RBC data are taken from Franco \textit{et al.}~\cite{franco2013abnormal}, diabetic-RBC data from Skovborg \textit{et al.}~\cite{SKOVBORG1966129}, and high-altitude cohorts with and without chronic mountain sickness (CMS) from Stauffer \textit{et al.}~\cite{stauffer2024making}. Both simulated curves---GD-RBC and CTR-RBC---are from~\cite{chai2025silico}; no new suspension-scale simulation was performed for the present study. All datasets exhibit pronounced shear-thinning behavior, with elevated viscosity at low shear rates reflecting enhanced cell--cell interactions, increased hematocrit, or impaired deformability. Blood from high-altitude individuals with CMS ($\eta \sim 70$--$90$~cP at $\dot\gamma \sim 1~\si{\per\second}$) shows markedly higher low-shear viscosity than non-CMS highlanders, consistent with hematocrit elevation ($H_t \approx 68\%$ vs.\ $57\%$). The normal-hematocrit curves (CTR-RBC, GD-RBC, diabetic RBC) lie within a narrow band at every sampled shear rate, whereas the high-hematocrit highlander cohorts remain well above them and approach that band only gradually as $\dot{\gamma}$ increases, consistent with progressive disruption of aggregates and flow-induced RBC alignment. Note the differing coverage of the datasets: the highlander cohorts extend to $\dot{\gamma} \approx 10^{2}~\si{\per\second}$ and the experimental CTR-, GD- and diabetic-RBC data to $\dot{\gamma} \approx 2.5\times10^{2}~\si{\per\second}$, while only the two simulated curves reach $\dot{\gamma} \sim 10^{3}~\si{\per\second}$.}
\label{fig:viscosity}
\end{figure}

To relate these thresholds to bulk flow, we placed the DPD suspension viscosity of control blood from our earlier Gaucher-disease study on the same platform~\cite{chai2025silico}, reproduced here for reference, alongside published rheograms for diabetic, Gaucher-disease, and high-altitude (CMS) blood (Fig~\ref{fig:viscosity}). All datasets exhibit pronounced shear thinning: $\eta$ decreased monotonically with $\dot\gamma$ over the interval each dataset covers---$\dot\gamma \in [1, 10^{3}]~\si{\per\second}$ for the two simulated curves, and narrower windows for the published experimental cohorts---and the magnitude in the low-shear regime varied by nearly an order of magnitude across the conditions compared, with the spread narrowing steeply as $\dot\gamma$ increased.

At $\dot{\gamma} \sim 1~\si{\per\second}$, CTR-RBC blood (hematocrit $H_t = 36\%$) yielded $\eta \approx 10$--$20$~cP, in line with standard hemorheological values~\cite{Chien1970Shear,SKOVBORG1966129,baskurt2003blood,nader2019blood}. In contrast, blood from reported CMS cohorts reached $\eta \approx 70$--$90$~cP at the same shear rate, corresponding to a $3$--$7\times$ increase relative to sea-level controls~\cite{stauffer2024making}. Published rheograms for type-2 diabetic RBCs~\cite{SKOVBORG1966129} and Gaucher-disease RBCs~\cite{franco2013abnormal} lie much closer to the control, within $\sim\!1.4\times$ of it (below); both simulated curves shown here---control blood and Gaucher-disease blood---come from our earlier study~\cite{chai2025silico}, whereas the diabetic and high-altitude curves are shown as published experimental data. Because resolving CMS-level hematocrits directly in particle simulation is computationally prohibitive, we instead assess the CMS elevation analytically, and find it to be dominated by the hematocrit contribution. Highlanders without CMS present a hematocrit near $57\%$, while CMS patients exceed $68\%$~\cite{stauffer2024making}, and most of the CMS elevation is already present in the highlanders without CMS: at $\dot\gamma = 2.25~\si{\per\second}$ the blood of the non-CMS cohort is $\sim\!4.4\times$ as viscous as the simulated control, and that of the CMS cohort only a further $1.5\times$, a margin that stays at $1.35$--$1.5\times$ over the shear rates measured (Fig~\ref{fig:viscosity}). For the $\sim\!11$-percentage-point rise in $\phi$ that separates the two cohorts, the Krieger--Dougherty-type scaling $\eta \sim (1-\phi/\phi_m)^{-[\eta]\phi_m}$~\cite{maron1956application,JAVADI2021hyperviscosity,javadi2022circulating} predicts at least a $2\times$ increase for an intrinsic viscosity $[\eta] = 2.5$ and any maximum packing fraction $\phi_m \leq 1$, and $\sim\!6\times$ for $\phi_m = 0.75$, more than the observed margin, so the hematocrit rise alone can account for the CMS excess. Mechanical heterogeneity between reticulocytes and mature RBCs---amplified at low shear by transient jamming and localized crowding---may add a second, smaller contribution, estimated at $\sim\!20$--$40\%$~\cite{nader2019blood,recktenwald2024morphology}, which these rheograms cannot separate from the hematocrit effect.

The three normal-hematocrit datasets---CTR, GD-RBC, and diabetic---lie within a narrow common band at every shear rate sampled, differing by no more than $\sim\!1.4\times$; the experimental curves extend only to $\dot{\gamma} \approx 2.5\times10^{2}~\si{\per\second}$, whereas the two simulated curves continue to $\dot{\gamma} \sim 10^{3}~\si{\per\second}$, where they reach $\eta \approx 3.7$--$4.5$~cP (Fig~\ref{fig:viscosity}). The high-hematocrit highlander cohorts are documented only up to $\dot{\gamma} \approx 10^{2}~\si{\per\second}$; over the window they share with the normal-hematocrit curves ($\dot{\gamma} \approx 2$--$90~\si{\per\second}$) the CMS excess viscosity falls from $\sim\!7\times$ to $\sim\!4\times$ that of CTR, so the gap narrows steeply but does not close within the measured range. Correspondingly, the shear-thinning ratio over that window is about $75\%$ greater for the CMS cohort ($\approx\!4.0\times$) than for CTR ($\approx\!2.3\times$), reflecting the stronger low-shear aggregation/crowding penalty. This steep narrowing toward a common high-shear limit reflects the well-known regime in which flow-induced membrane alignment and deformation suppress aggregation and crowding, and even moderately stiffer cells are forced to deform sufficiently to reduce effective flow resistance~\cite{Chien1970Shear,baskurt2009guidelines}. The overall picture is therefore that mechanical heterogeneity and hematocrit exert their strongest effects under the low-shear conditions relevant to post-capillary and splenic microcirculation, while high-shear arterial flow progressively suppresses them---an interpretation compatible with the single-cell mechanisms identified in the confined-flow simulations, where deformability differences and single-file queuing likewise decide whether a cell passes only near the threshold driving and, well above it, set a delay rather than retention.

\subsection*{Linking reticulocyte dynamics in the spleen to altitude-related clinical outcomes}

\begin{figure}[!tbp]
\begin{center}
\includegraphics[width=1.00\textwidth]{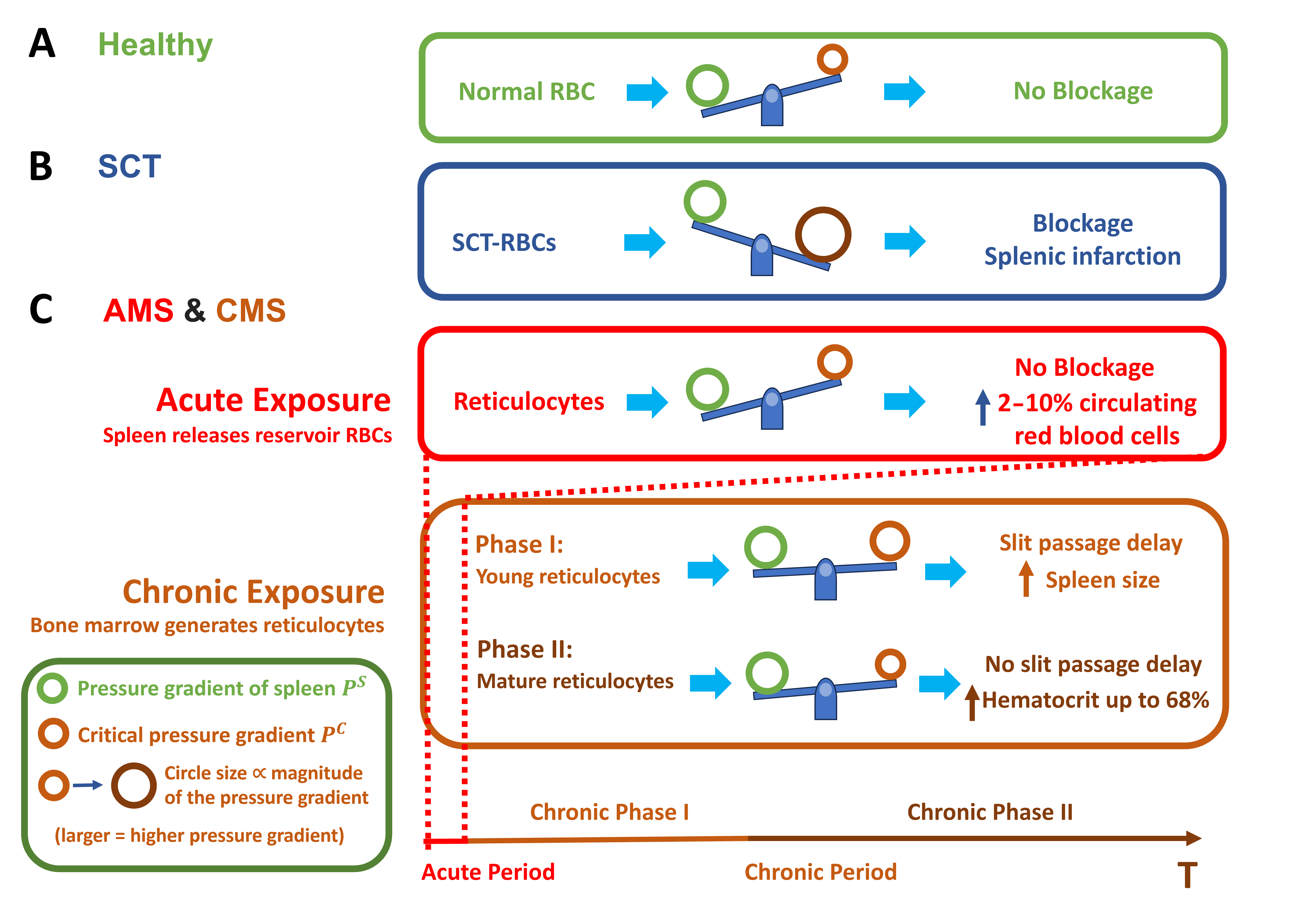}
\end{center}
\vspace{-0.1in}
\caption{\small{\bf Conceptual framework linking splenic pressure gradients, reticulocyte maturation, and SCT-RBC behavior under acute and chronic altitude exposure.}
Throughout, the balance depicts slit passage as a competition between two pressure gradients: the splenic pressure gradient $P^{S}$ (green, left pan) that drives a cell through the slit, and the cell's own critical pressure gradient $P^{C}$ (orange, right pan; identical to the $\Delta P_c$ used throughout the text) that it must overcome. Circle size is proportional to the magnitude of each gradient, and the beam tips toward the larger one: the cell clears the slit when $P^{C} < P^{S}$ (beam down on the green side) and is retained when $P^{C} > P^{S}$ (beam down on the orange side).
(A) Healthy. A mature discocyte has a low $P^{C}$ that stays well below $P^{S}$, so it passes without blockage---the baseline against which the other two conditions are read.
(B) Sickle cell trait. Under profound hypoxia a subpopulation of SCT-RBCs stiffens enough that $P^{C}$ approaches or exceeds $P^{S}$ (largest orange circle); these cells become mechanically trapped at the slit, producing blockage and acute splenic infarction.
(C) Acute and chronic mountain sickness. \textit{Acute exposure} (red): transient hypoxic stress induces splenic contraction and release of reservoir RBCs; reticulocytes have a $P^{C}$ only modestly above that of mature cells and well below $P^{S}$, traverse the slits without blockage, and transiently elevate the circulating RBC count by $\sim$2--10\%. \textit{Chronic exposure} (brown): sustained hypoxia activates bone-marrow erythropoiesis and progressive reticulocyte production. In Phase~I, young R1-like reticulocytes have the highest shear modulus of the reticulocyte spectrum and a $P^{C}$ closest to, but still below, $P^{S}$ (beam tipped only slightly toward the green side); they pass the slits $\sim\!30\%$ more slowly than mature cells (Fig~\ref{fig:microchannel}D), contributing to transient splenic fullness. In Phase~II, they mature toward R2/R3 and then discocyte morphology, lowering $P^{C}$, restoring efficient passage, and driving the hematocrit elevation characteristic of CMS. The timeline at the bottom places the acute response and the two chronic phases on a common time axis $T$.}
\label{fig:conceptual}
\end{figure}

We synthesized the preceding results into a mechanistic framework in which reticulocyte mechanics shape the hematological phenotype of high-altitude exposure (Fig~\ref{fig:conceptual}). The framework distinguishes an acute and a chronic exposure regime, and connects single-cell deformability, single-file crowding, and splenic filtration to population-level outcomes observed in AMS, CMS, and SCT-related splenic syndrome.

\textit{Acute exposure.} A rapid ascent or other acute hypoxic stress induces sympathetically driven splenic contraction and partial emptying of the splenic reservoir, transiently raising circulating RBC count by $\sim$2--10\%~\cite{stewart2002spleen} (Fig~\ref{fig:conceptual}C, acute-exposure box). Reticulocytes in the released population carry a $\Delta P_c$ only $20$--$31\%$ above that of the mature discocyte and far below that of the sickled cell (Fig~\ref{fig:dpc_ladder}B), and every reticulocyte model cleared the IES-like slit at the loading simulated here (Fig~\ref{fig:microchannel}D); they are therefore expected to traverse the splenic IES without substantial retention. In SCT individuals, by contrast, deoxygenation at altitude partially stiffens a subpopulation of HbAS erythrocytes whose $\Delta P_c$ then approaches or exceeds the prevailing splenic trans-slit pressure, leading to mechanical trapping, localized red-pulp ischemia, and the clinical syndrome of acute splenic infarction at altitude~\cite{goodman2014splenic}. The acute response is thus a balance between reticulocyte augmentation of circulating RBCs and preferential sequestration of mechanically compromised cells.

\textit{Does splenic emptying protect SCT carriers?} The acute response contains an apparent tension worth making explicit: if hypoxia contracts the spleen and expels part of its reservoir, the organ holds fewer cells, and one might expect SCT-RBCs to be correspondingly \emph{less} likely to obstruct it. The two processes, however, act on different populations and in opposite directions. Splenic contraction expels cells already resident in the sinuses and venous outflow---cells that have by definition passed the filtration bed---whereas splenic syndrome is caused by cells that cannot traverse the IES at all. Emptying the reservoir does not remove that barrier; by driving a larger flux of cells through the cords per unit time, contraction increases the rate at which stiffened HbAS cells are presented to the slits, so even an unchanged per-cell retention probability yields more retention events per unit time. Contraction also reduces cord volume and thereby raises the local cell concentration in the filtration bed, and our pairwise results show that single-file congestion \emph{raises} rather than lowers the passage threshold of the cell behind. Acute splenic emptying is therefore not protective in SCT: it increases both the challenge rate and the local crowding at precisely the moment when deoxygenation is stiffening the cells, which is consistent with the clinical observation that splenic syndrome occurs during rapid ascent rather than after sustained residence at altitude~\cite{goodman2014splenic}.

\textit{Chronic exposure.} Sustained hypoxemia drives enhanced erythropoiesis and persistent reticulocytosis~\cite{siebenmann2017regulation} (Fig~\ref{fig:conceptual}C, chronic-exposure box). We decompose this adaptation into two stages. In Phase~I, an initial surge of the most immature, R1-like reticulocytes enters the circulation; these cells have the highest shear modulus of the reticulocyte spectrum, so they cross the IES-like slit $\sim\!30\%$ later than the control discocyte (Fig~\ref{fig:microchannel}D), carry the highest critical driving of the reticulocyte subtypes (Fig~\ref{fig:dpc_ladder}B), and are the most likely of them to be transiently retained, contributing to splenic fullness during early acclimatization~\cite{moreau2023physical}. In Phase~II, reticulocytes mature in the circulation: excess membrane is shed through vesiculation/pitting and the spectrin cytoskeleton is remodeled, shifting the population toward R2/R3 and then mature discocytes~\cite{chasis1989membrane,mohandas2008red,Renoux2019Impact}. More mature reticulocytes traverse the spleen without retention and feed a progressively expanding circulating RBC pool. The endpoint is the characteristic CMS phenotype: hematocrit 60--70\%~\cite{villafuerte2022highaltitude}, low-shear viscosity 70--90~cP (Fig~\ref{fig:viscosity}), and elevated microvascular resistance that is mitigated---but not eliminated---by the low single-cell $\Delta P_c$ of the maturing reticulocyte population, even as single-file crowding at elevated hematocrit works against transit.

The timeline at the bottom of Fig~\ref{fig:conceptual}C summarizes the temporal progression from acute splenic release and early reticulocyte influx to chronic erythropoietic amplification. Together, these components link cellular-scale reticulocyte mechanics---deformability, crowding-limited transport, and splenic-transit thresholds---to organism-level hematological outcomes observed during acute and chronic high-altitude exposure, and articulate a clean biophysical rationale for why SCT individuals and CMS patients lie at opposite ends of the same mechanical axis.

\section*{Discussion}

Using a combined experimental--computational framework, this study elucidates how reticulocyte morphology, deformability, and cell--cell interactions govern RBC transport and splenic retention under microconfinement, and connects these mechanisms to the hematological phenotypes of acute and chronic mountain sickness. Single-cell and pairwise simulations on one DPD platform, read against suspension rheology, place benign AMS acclimatization, CMS hyperviscosity, and SCT splenic syndrome on a shared mechanical axis---the critical pressure gradient $\Delta P_c$ relative to the control-discocyte threshold. Two results anchor this axis. First, a $5~\si{\micro\metre}$ microchannel registers the stiffness of a reticulocyte only weakly (R1 takes at most $\sim\!18\%$ longer than CTR to cross it), whereas a $1.5~\si{\micro\metre}$ splenic slit delays R1 by $\sim\!30\%$, both geometries ordering the models by shear modulus. Measured with each geometry at its own loading rather than at a common gradient, this places the splenic slit, rather than the capillary-scale constriction, as the stage at which reticulocyte stiffness matters~\cite{safeukui2018sensing,qiang2023microfluidic}, while whether a cell passes the slit at all remains set by its surface area and volume~\cite{pivkin2016biomechanics,dao2021erythrocyte}. Second, pairwise simulations resolve the cooperative question with a nuanced answer. A leading cell never lets a follower pass a sub-cellular pore below its own single-cell threshold---across the driving forces and realizations swept in this geometry no leader converted a retained follower into a passing one---and in a single-file slit a leader in fact \emph{raises} the follower's threshold, because the follower queues behind it: it reaches the slit as the leader leaves and then crosses in about its lone transit time. The order-of-magnitude reduction that a wake-``unjamming'' mechanism would suggest is therefore absent, and a cell passes most easily alone. What the leader's compliance does control is how long the follower waits: in the physiologically scaled slit, the trailing cell required a critical driving $\sim\!5\%$ higher behind a leader $1.3\times$ its own shear modulus than behind one at $0.6\times$, because the compliant leader deforms through and vacates the pore sooner. In crowded single-file flow, a cell's deformability therefore acts on the cells behind it through this leader-compliance effect---not an absolute rescue---layered on top of the single-cell $\Delta P_c$ that governs whether each cell passes.

Across cell populations, the single-cell critical driving $\Delta P_c$ to initiate slit passage rose by more than $2.15\times$ across the CTR/R1--R3/sickled panel, measured in a single geometry and quoted relative to the CTR discocyte---the reticulocyte subtypes $20$--$31\%$ above it, the sickled cell more than $2.15\times$ it, a threshold the ladder bounds but does not locate. Importantly, at any splenic operating pressure lying above the reticulocyte thresholds but below the sickled-cell threshold, the spleen passes control discocytes, reticulocytes and unsickled HbAS cells while retaining the cells that have sickled, and this provides a mechanical rationale for splenic syndrome in SCT individuals on rapid altitude ascent, a phenomenon that has been documented clinically for decades but has lacked a clean single-cell-to-population explanation~\cite{goodman2014splenic}. In the same framework, the R1--R3 reticulocytes, although stiffer than mature discocytes, transit the splenic IES with $\Delta P_c$ well below the sickled-cell threshold, consistent with their release without retention during splenic contraction.

Bulk rheology links microscale deformation to macroscale flow resistance. Across all conditions examined, blood exhibited strong shear thinning. For the normal-hematocrit curves, $\eta$ decreased from $\sim\!10$--$20$~cP at $\dot{\gamma} \sim 1~\si{\per\second}$ to $\sim\!3.7$--$4.5$~cP at $\dot{\gamma} \sim 10^{3}~\si{\per\second}$, a $\sim\!4\times$ reduction; the CMS cohort sits far higher at low shear ($\sim\!70$--$90$~cP at $\dot{\gamma} \sim 1~\si{\per\second}$) and, over the $\dot{\gamma} \lesssim 10^{2}~\si{\per\second}$ window in which it was measured, thins more steeply than CTR ($\approx\!4.0\times$ versus $\approx\!2.3\times$)---both behaviors consistent with the classical RBC rheograms of Chien and co-workers~\cite{Chien1970Shear}. The CMS elevation at low shear is dominated by hematocrit (68\% vs.\ 57\% in non-CMS highlanders~\cite{stauffer2024making}); mechanical heterogeneity may add a second, smaller contribution ($\sim\!20$--$40\%$ in published estimates~\cite{nader2019blood,recktenwald2024morphology}), which the rheograms of Fig~\ref{fig:viscosity} cannot separate from the hematocrit effect. We expect this contribution to follow the same single-cell deformability axis that sets $\Delta P_c$ at the microscale: reticulocyte-enriched populations, which sit between control discocytes and sickled cells on this axis, should deform more readily under low-shear crowding than the stiffest populations, tempering the bulk-level viscosity impact of cell-population stiffening, whereas increasing shear should progressively suppress the difference as flow-induced alignment sets in; suspension simulations of mixed populations, which we did not perform, would test this.

The conceptual framework of Fig~\ref{fig:conceptual} synthesizes these findings into a unified picture of reticulocyte mechanics at altitude. In AMS, splenic contraction releases a reticulocyte pool whose $\Delta P_c$, only modestly above that of mature cells, allows retention-free passage and accounts for the observed $\sim$2--10\% transient increase in circulating RBCs~\cite{stewart2002spleen}. In SCT, the opposite limit is reached under profound hypoxia: a subpopulation of HbAS erythrocytes partially sickles and stiffens enough to meet or exceed the prevailing splenic trans-slit pressure, infarcting the red pulp and explaining the ``splenic syndrome'' of mountaineers~\cite{goodman2014splenic}. In CMS, sustained reticulocytosis under chronic hypoxia and the ensuing hematocrit elevation to 60--70\% drive the low-shear hyperviscosity observed in the Stauffer \textit{et al.} cohorts and estimated analytically above, while the continued presence of reticulocytes---whose $\Delta P_c$ sits well below that of sickled cells---should temper the bulk suspension viscosity relative to a stiffer population, even as single-file crowding at the elevated hematocrit works against microvascular transit~\cite{stauffer2024making}. Several prior studies on high-altitude adaptation have emphasized the role of hemoglobin oxygen affinity~\cite{storz2016hemoglobin}, plasma-volume regulation~\cite{stembridge2019plasma}, and genetic factors in Andean and Tibetan populations~\cite{beall2007two,simonson2010genetic}; our framework is complementary rather than competing, since it isolates the mechanical component of the adaptation---a component that becomes dominant in the microcirculation and in the spleen.

Several limitations motivate future work. First, the present DPD framework captures mechanical and hydrodynamic effects but does not explicitly include biochemical regulation, membrane remodeling during reticulocyte maturation, or macrophage-mediated interactions in the splenic red pulp; coupling to explicit lipid/protein membrane models (e.g.\ OpenRBC~\cite{tang2017openrbc,li2021How} or two-component spectrin-lipid membranes~\cite{li2014erythrocyte,wang2024two}) and to the recently developed multiscale signaling--biophysical pipeline for macrophage-mediated RBC clearance in sickle cell and Gaucher disease~\cite{chai2026multiscale} would be a natural next step. Second, while the pairwise simulations give a parameter-resolved picture of single-file crowding, fully developed many-body dynamics at physiological hematocrit (up to 60--70\% in CMS) and across spatially heterogeneous splenic geometries remain computationally expensive; the emergent $\Delta P_c$ reported here should be validated against larger-scale simulations and against microfluidic cohorts with mixed reticulocyte and SCT cell populations~\cite{qiang2023microfluidic,kumari2024measuring,man2020microfluidic}; covering that space at this cost will benefit from automated discovery and tuning of the numerical schemes themselves~\cite{toscano2026graftathena}. In particular, the intermittent, bursty passage of red cells observed at splenic inter-endothelial slits in vivo was not reproduced by our short pairwise queues under constant pressure, where transit is threshold-gated cell by cell; capturing such collective ``stick--slip'' release plausibly requires both a much longer upstream queue---so that back-pressure can accumulate and discharge as an avalanche at physiological hematocrit---and a time-varying, pulsatile trans-slit pressure, which we identify as targets for future larger-scale and pulsatile-driving simulations. Third, our parameterization of the sickled cell and of the reticulocyte subtypes rests on published ektacytometry and microfluidic data rather than on donor-matched measurements, and the fraction of an SCT carrier's cells that sickle at a given altitude is not known, so the model places the sickled cell on the mechanical axis but cannot by itself predict what fraction of that carrier's cells the spleen will retain; targeted experiments on AMS and CMS cohorts, and on SCT individuals before and after altitude exposure, would sharpen the quantitative predictions made here, as patient-derived parameterization has done for other red-cell disorders~\cite{chai2026gaucher}. Relatedly, the experimental material analyzed here derives from HbSS donors rather than from healthy or altitude-exposed individuals, so the morphological class proportions reported above should be re-established in a healthy cohort before they are read as normal reticulocyte demographics; the mechanical descriptors are unaffected, because they were assigned from the reported mechanics of reticulocytes rather than from class counts. Fourth, supravital staining of an additional donor sample shows that reticulum is concentrated in the multilobular class but that most cup-shaped cells in that sample are stomatocytes, indistinguishable from cup-shaped reticulocytes in bright field; the stain cannot establish the cup-shaped reticulocyte stage itself, whose identity here rests on the maturation literature, because NMB does not reliably separate the few discrete granules of a late reticulocyte from Howell--Jolly or Heinz bodies. Morphology-based sorting of reticulocyte subtypes should therefore be validated against a molecular marker of immaturity---definitively, by sorting on an RNA dye and CD71 and imaging the sorted fractions~\cite{Leduc2025Comprehensive}---in any cohort where the proportions of the subtypes, rather than their mechanics, are the quantity of interest. Fifth, the confined-flow simulations are driven by a uniform body force in a streamwise-periodic box, which compresses the DPD solvent ahead of each constriction and rarefies it behind, a density ramp that can still be building while the cells cross. Because every cell type within a comparison is released from the same position at the same instant, the orderings and ratios reported here are unaffected to first order, although absolute flow rates and passage times are specific to this driving protocol; simulations with shorter channels or a stiffer solvent equation of state, which would remove the ramp, are a natural next step.

In summary, the results highlight how distinct vascular compartments impose different mechanical selection pressures. A capillary-scale constriction registers the stiffness of a reticulocyte only weakly, whereas a splenic slit delays it according to its shear modulus and retains only the most mechanically compromised cells; reticulocytes, whose $\Delta P_c$ remains far below that of the mechanically compromised cells the spleen removes, clear these constrictions and sustain their own passage, although in single-file crowding a cell lodged in a pore can still block those queued behind it. This is particularly relevant in hypoxic conditions, where splenic contraction releases reticulocytes into circulation while rigid cells may approach or exceed the available splenic pressure gradient and become trapped. Under chronic hypoxia, sustained reticulocytosis links cellular-scale mechanics to splenic filtration, hematocrit elevation, and hyperviscosity. Overall, this work establishes mechanical heterogeneity, and particularly the position reticulocytes occupy on the deformability axis, as an active and central participant in confined blood flow and collective transport, with direct implications for altitude medicine, SCT-related splenic syndrome, and the broader class of red-cell disorders in which cell-population composition sets microcirculatory performance.

\section*{Materials and methods}

\subsection*{Ethics statement}
Whole HbSS blood samples were obtained from Boston Children's Hospital and Massachusetts General Hospital under excess human material protocols approved by the respective institutional review boards, the Boston Children's Hospital Institutional Review Board and the Partners HealthCare Institutional Review Board, with a waiver of consent. Blood samples were then promptly transported to MIT for experiments under an approved exempt protocol (Massachusetts Institute of Technology IRB protocol E-1523).

\subsection*{Computational framework}

We use a multiscale dissipative particle dynamics (DPD) framework to resolve the biomechanics and confined transport of single RBCs and cell pairs---control discocytes, reticulocytes and sickled cells; the suspension viscosity shown for comparison was computed on the same platform in earlier work~\cite{chai2025silico}. DPD represents a coarse-grained group of atoms as a single bead interacting pairwise through conservative, dissipative, and random forces, the latter two tied by the fluctuation--dissipation theorem~\cite{hoogerbrugge1992simulating,groot1997dissipative}. This preserves hydrodynamic interactions while remaining computationally tractable for RBC suspensions at physiological hematocrit~\cite{fedosov2010multiscale,Fedosov2011Multiscale,Fedosov2011Predicting,chang2016md}. Full DPD force definitions, parameter values, and unit mappings are collected in S1 Text, Section~1.

\subsection*{Dissipative particle dynamics (DPD) model}

Blood plasma and the RBC cytoplasm are both represented by DPD particles interacting through the standard soft conservative, dissipative, and random forces~\cite{groot1997dissipative,hoogerbrugge1992simulating,Fedosov2011Multiscale}. The plasma viscosity at 37~\si{\celsius} ($\eta_p \approx 1.2$~cP) was calibrated by reverse Poiseuille flow~\cite{Fedosov2011Predicting}, and the cytoplasm viscosity is set equal to that of plasma; this simplification has been shown to reproduce RBC suspension rheology quantitatively in the physiological shear-rate range~\cite{fedosov2010multiscale,Fedosov2011Predicting}. Membrane vertices interact with the solvent and with the frozen wall particles through the same pair parameters as solvent particles with one another, except for the posts that form the elongated slit of the pairwise and critical-driving assays and the walls of the microchannel and the tube, whose interaction with membrane vertices has a stronger repulsion and dissipation and a random force corresponding to $1.5$ times the solvent temperature; all values are listed in Table~S1 in S1 Text. The pairwise assay adds a short-range repulsive contact term between membranes and between membranes and the posts (below).

Boundary conditions are problem-specific: in the suspension-rheology simulations of~\cite{chai2025silico}, the domain was periodic in the flow and vorticity directions, with no-slip walls moving at constant velocity along the gradient direction; in the confined-flow simulations of this study, a constant pressure gradient drives RBCs through capillary-like channels or splenic-slit geometries. How each assay was initialized is stated below.

\subsection*{RBC model}

Each RBC is represented as a triangulated surface of $N_v = 500$ vertices connected by $N_s$ springs with dashpots, following the multiscale coarse-grained formulation of~\cite{fedosov2010multiscale,Fedosov2011Multiscale,chai2026quantifying,chang2016md,li2014erythrocyte,li2016modeling}. Membrane elasticity is captured by a combined worm-like-chain (WLC) and power-function (POW) bond potential, while local area, total area, and enclosed-volume constraints enforce near-incompressibility (see S1 Text, Section~2)~\cite{skalak1973strain,helfrich1973elastic}. Resistance to out-of-plane deformation is modeled by a dihedral-angle bending potential~\cite{fedosov2010multiscale}. The resulting macroscopic shear modulus $\mu$ and bending modulus $k_c$ are obtained from the model parameters through the closed-form mappings derived in~\cite{fedosov2010multiscale}, where this membrane model was validated against optical-tweezer stretching of healthy RBCs~\cite{suresh2005connections}; stretching runs of our implementation show that the realized shear modulus scales with its target value (S1 Text, Section~4).

For the mature control discocyte (CTR-RBC) we use the standard values $\mu_{0} = 4.73$~\si{\micro\newton\per\metre}, $k_{c,0} = 2.4 \times 10^{-19}$~J, and $A_{0} = 132.9$~\si{\micro\metre\squared}, $V_{0} = 92.5$~fL, consistent with classical experimental measurements on deoxygenated and oxygenated human erythrocytes~\cite{evans1973new,fedosov2010multiscale,suresh2005connections}. The suspension-viscosity curves of Fig~\ref{fig:viscosity} are reproduced from~\cite{chai2025silico}, which describes the cell--cell interaction potentials used to obtain them.

\subsection*{Parameter setup for reticulocyte models}

Reticulocytes differ from mature discocytes in three coupled biophysical dimensions that we parameterize independently. First, they retain excess plasma-membrane area: quantitative vesiculation and pitting experiments indicate that $\sim$15--30\% of the membrane is shed during reticulocyte maturation~\cite{chasis1989membrane,malleret2013significant,mohandas2008red}. Second, their spectrin--actin cytoskeleton is only partially remodeled, yielding an effective membrane shear modulus that exceeds the mature-discocyte value by factors of $\sim$1.3--1.8 as measured by ektacytometry and microfluidic deformability assays~\cite{cluitmans2012red,Renoux2019Impact,kumari2024measuring}. Third, residual organelles and a lipid composition that is not yet fully restructured modify the bending rigidity and local curvature elasticity~\cite{chasis1989membrane,moreau2023physical}. Guided by these measurements, we construct three reticulocyte models (R1, R2, R3) that span the observed multilobular-to-near-discocytic maturation axis (Table~\ref{tab:mechanical_properties}):
R1 (multilobular, early reticulocyte) has the largest excess area ($A_0 = 160~\si{\micro\metre\squared}$, $S/V = 1.56~\si{\per\micro\metre}$), the highest shear modulus ($\mu = 8.28~\si{\micro\newton\per\metre}$), and a moderately elevated bending modulus ($k_c = 4.8 \times 10^{-19}$~J), representing an unremodeled cytoskeleton and residual membrane--cytoskeletal couplings. R2 (cup-shaped, intermediate) has reduced excess area ($A_0 = 150~\si{\micro\metre\squared}$, $S/V = 1.52~\si{\per\micro\metre}$), a lower shear modulus ($\mu = 7.00~\si{\micro\newton\per\metre}$), and the same bending modulus as R1. R3 (near-discocytic, late reticulocyte) has the smallest excess area among the reticulocyte set ($A_0 = 142~\si{\micro\metre\squared}$, $S/V = 1.48~\si{\per\micro\metre}$) and the lowest shear modulus ($\mu = 6.29~\si{\micro\newton\per\metre}$), but the highest bending rigidity ($k_c = 7.2 \times 10^{-19}$~J), to capture localized curvature-stiffening from residual vesicular and organellar patches~\cite{moreau2023physical,malleret2013significant}.
\noindent
\textit{Shape models of Fig~\ref{fig:morphology}A.} The membrane configurations shown beneath the bright-field images were generated separately from the R1--R3 flow models, and serve only to connect a two-dimensional silhouette to a membrane area. Starting from the mature-discocyte mesh ($500$ vertices, $996$ triangular faces, $A_0 = 132.87~\si{\micro\metre\squared}$, $V_0 = 92.45$~fL), the enclosed-volume target of the area/volume potential was lowered at fixed membrane area and a spontaneous dihedral angle of $60^{\circ}$ imposed, and the mesh relaxed for $10^{4}$ steps; this produces the cup morphology at $A = 132.98~\si{\micro\metre\squared}$, $V = 75.47$~fL ($S/V = 1.76~\si{\per\micro\metre}$, sphericity $0.650$). The relaxed meshes were then checked, without any further adjustment, against a separate census of the two-dimensional footprints that individual suspended cells present in a wet mount ($n = 184$ scorable cells of $210$ segmented, $0.0575~\si{\micro\metre}$ per pixel): the discocyte mesh projects $47.6~\si{\micro\metre\squared}$ face-on, the orientation at which a discocyte settles on a slide, against a measured median of $43.9~\si{\micro\metre\squared}$ ($-8\%$), and the cup mesh projects $30.2~\si{\micro\metre\squared}$ averaged over orientation against a measured median of $33.7~\si{\micro\metre\squared}$ ($+11\%$). That wet-mount footprint is a different observable from the projected surface area of Fig~\ref{fig:morphology}B and orders the three classes in the opposite direction, because a lobed or cupped cell folds a large membrane into a compact silhouette whereas a flatter discocyte presents its membrane nearly face-on; the two are used for different purposes here and are never compared directly. The areas plotted in Fig~\ref{fig:morphology}B are apparent projected areas from bright-field images of a separate preparation, and enter the argument only through the $17\%$ decrease they show across the maturation sequence; they are not used as absolute areas. Taken at face value they lie above the bound $A_{\rm proj} \leq S/2$ that any closed surface satisfies, evaluated with the reference areas of Table~\ref{tab:mechanical_properties} ($80$, $75$ and $71~\si{\micro\metre\squared}$), by $12\%$, $12\%$ and $5\%$ for the multilobular, cup-shaped and discocyte classes---an offset in the same direction for all three, which leaves their ordering unaffected. Averaged over orientation, the projected area of the discocyte and cup meshes is $0.245$ and $0.227$ of their membrane areas, below the quarter that Cauchy's theorem gives for a convex body because both shapes are concave, and the largest single projection reaches only $0.36$ of the membrane area, well inside the bound $A_{\rm proj} \leq S/2$. These shape models carry a lower reduced volume than the R1--R3 flow models of Table~\ref{tab:mechanical_properties}, which are parameterized from reported membrane-loss fractions rather than from a target morphology: a cup deep enough to be recognizable in a micrograph requires a reduced volume below what the R1--R3 excess-area assignment provides. The two sets are used for different purposes and are not interchangeable.

R1 is therefore the stiffest of the three in shear, and carries the largest excess membrane area, but shares the lowest bending modulus with R2; R3 is the softest in shear and the stiffest in bending. All three descriptors enter the shape-transition simulations (Fig~\ref{fig:shape_transition}; below). The transport simulations vary the shear modulus alone, and there the excess areas and bending moduli listed here enter only the tests of \nameref{S1_Text}, Sections~5 and~7. The parameter set was fixed before the flow simulations and was not adjusted afterwards.

\subsection*{Experiment setup}

\textit{Preparation of RBC samples}

Whole blood samples were collected from sickle cell anemia (HbSS) patients at Massachusetts General Hospital (see \textit{Ethics statement}). Following a pretreatment procedure previously described by our group \cite{du2015kinetics}, HbSS samples with heterogeneous RBC types were gently washed three times with phosphate-buffered saline (1$\times$ PBS; CaCl$_2$-free, MgCl$_2$-free; pH 7.4; Gibco) by centrifugation at 1,500 rpm for 3 min at room temperature. The washed RBC pellets were then resuspended in PBS containing 1\% (w/v) bovine serum albumin (BSA; EMD Millipore) to achieve a hematocrit of 2\%. Samples were stored at 4~\si{\celsius} and used within 24~h.

\noindent
\textit{Microfluidic channels}

The microfluidic device was fabricated by bonding polydimethylsiloxane (PDMS) to a glass slide following previously reported methods \cite{qiang2019mechanical}. A customized SU-8 mold was used to cast a degassed PDMS mixture (base:curing agent = 10:1, w/w). The channel dimensions were $5\times5\times50$~\si{\micro\metre}. Prepared RBC samples were introduced into the microfluidic device using a water column at an approximate flow velocity of 350~\si{\micro\metre}/min. Bright-field images and videos of RBCs were acquired using a high-resolution CMOS camera (The Imaging Source, Charlotte, NC, USA) mounted on an Olympus IX71 inverted microscope (Olympus America, Breinigsville, PA, USA), equipped with a 60$\times$ oil-immersion objective lens (NA = 1.25).

\noindent
\textit{Supravital staining of reticulocytes}

To establish which of the bright-field morphological classes actually retain ribosomal RNA, a sample from an additional HbSS donor was stained supravitally with New Methylene Blue (NMB), the classical reticulocyte stain~\cite{brecher1949new}, and imaged unfixed as a wet mount with the same microscope and $60\times$ objective as the microfluidic experiments.
 Two preparations of differing staining strength were acquired, comprising $13$ and $15$ independent fields respectively. Cells were segmented individually by placing the contour on the dark membrane rim rather than on the surrounding diffraction halo, and staining was quantified as per-channel optical density referenced to each field's own background, which removes the effect of the differing white balance between the two preparations.

In the more weakly stained preparation, which supplies every class-conditional rate reported here, cell shape and staining were read in two separate passes, so that morphology was never classified with the stain in view; the more strongly stained preparation was used only for a targeted survey of stained cells, in which shape was recorded with the stain visible. NMB is a true chromogen that absorbs green light and transmits blue: measured over $42$ confirmed granules, the optical-density excess of a granule over the surrounding cytoplasm was $+0.082$ in the green channel, $+0.041$ in the red and $-0.006$ in the blue. Morphology was therefore scored from blue-channel renderings, in which the reticulum is invisible while cell shape is fully preserved, and the presence of stain was adjudicated afterwards from the color images presented in a different, shuffled order. In total $210$ cells were classified by shape and $139$ of these were additionally adjudicated for stain; reported population fractions are re-weighted for the stratified sampling used to select them. Staining was scored visually, as in the reference manual method, because at this staining intensity automated detection is confounded by chromatic fringing along the internal slit of a stomatocyte, which mimics a stain granule. The pixel size was taken as $0.0575~\si{\micro\metre}$ ($60\times$ objective on a $3.45~\si{\micro\metre}$ sensor pitch), which places the membrane rim of round cells at a diameter of $7.4~\si{\micro\metre}$; the resolution tag written by the slide scanner is inconsistent with this and was not used.

\subsection*{Simulation setup}

Each simulation class was specified by defining the computational domain, boundary conditions, particle resolution, and flow-driving mechanism, as summarized below and tabulated in Table~S4 in S1 Text.

\textit{Flow-induced shape transition.} The shape response of Fig~\ref{fig:shape_transition} was computed for a single cell in a straight cylindrical channel of diameter $D = 9~\si{\micro\metre}$, periodic along its axis with a period of $29.07~\si{\micro\metre}$, which sets the volume fraction to $C = 0.05$. The channel wall is the block of frozen DPD particles that fills the box outside the lumen, and the solvent--wall and membrane--wall interactions are those of the microchannel below (Table~S1 in S1 Text). Each reticulocyte model carried the three descriptors of Table~\ref{tab:mechanical_properties}: its shear modulus; its $S/V$, imposed as in the surface-to-volume tests of the slit (below) by lowering the reference volume at the control membrane area to $V_0 = 92.46 \times 1.44/(S/V)$~fL ($85.35$, $87.59$ and $89.96$~fL for R1, R2 and R3); and its bending modulus, set by scaling the bending coefficient of the membrane model by $k_c/k_{c,\mathrm{CTR}}$ ($2\times$ for R1 and R2, $3\times$ for R3; \nameref{S1_Text}, Section~2). The cell was relaxed without driving for $2 \times 10^{4}$ steps, in which the enclosed volume settles at its new reference value ($10^{4}$ steps for the control, whose reference values are unchanged). The flow was then driven by a uniform body force $x_f$ on every solvent and membrane particle, from $x_f = 0.2$ to $3.2$ in DPD units, with runs at $x_f = 0$ as the zero-flow reference, in three independent realizations of each cell type at each driving and in three more at $x_f \geq 0.5$ (up to $x_f = 2.5$ for the control). Runs lasted $1.5 \times 10^{5}$ steps, lengthened up to $9.3 \times 10^{5}$ steps at the weakest drivings; every reticulocyte run at $x_f \geq 0.8$ lasted $4.5 \times 10^{5}$ steps, as did the control runs at $x_f = 1.0$--$1.6$, where shorter runs had ended while the cell was still changing shape. Every run whose $\lambda_3$ over the last quarter differed from that over the third quarter by more than $0.15~\si{\micro\metre\squared}$ (16 of the runs here and 12 of the shear-modulus-only series below, most of them near the transition, where a cell can remain in an intermediate shape for much of a run before it turns into a parachute) was repeated with the same input and random seed for at least twice its length and at least $9 \times 10^{5}$ steps, and the longer run replaced it; every run then met this criterion. $\bar V_x$, the mean solvent velocity along the axis, and $\lambda_3$, sampled every $2000$ steps, are averages over the second half of each run, and $\lambda_3^{\mathrm{eq}}$ is the mean over the three zero-flow realizations of the same cell type. For each realization, $\bar V_x^{c}$ is the velocity at which $\Delta\lambda_3$ crosses midway between its mean over $x_f = 0.5$ and $0.63$ ($\bar V_x \approx 60$--$100~\si{\micro\metre}/\mathrm{s}$, the lowest drivings that all six realizations share, below the transition in all but one R3 realization at each) and its mean between $480$ and $1000~\si{\micro\metre}/\mathrm{s}$, interpolated linearly in $\log \bar V_x$; each reticulocyte model was compared with the control by Welch's two-sided $t$-test over the six realizations. The next driving, $x_f = 0.8$ ($\bar V_x \approx 150~\si{\micro\metre}/\mathrm{s}$), is already inside the transition of R3 and is therefore not part of the lower level; including it gives $265 \pm 37$, $162 \pm 19$, $211 \pm 43$ and $207 \pm 58~\si{\micro\metre}/\mathrm{s}$ for the control, R3, R2 and R1 ($p = 0.0004$, $0.04$ and $0.07$) and leaves the conclusions unchanged. The shear-modulus-only series of \nameref{S1_Text}, Section~8, follows the protocol of the control, with every model at the control reference volume and bending modulus.

\textit{Microchannel constriction.} Microfluidic constriction assays were modeled using a rectangular channel of $60 \times 10 \times 10~\si{\micro\metre\cubed}$ containing a narrow constriction $30~\si{\micro\metre}$ long, $5~\si{\micro\metre}$ wide, and $2.7~\si{\micro\metre}$ deep. This constriction shares its $5~\si{\micro\metre}$ width with the PDMS device above but not its depth or length ($5$ and $50~\si{\micro\metre}$ there): it reproduces the transverse confinement of the experiment rather than the device itself, and the imaged and simulated cells of Fig~\ref{fig:platform}C are set side by side for shape, not for speed. The flow was driven by a uniform body force $x_f$ on every solvent and membrane particle, swept over $x_f = 0.8$, $1.0$, $1.2$, $1.4$, $1.7$ and $2.0$ in DPD units, with three independent realizations of each cell type at each driving; R1 at $x_f = 0.8$ was run six times. Two runs ended in a numerical membrane rupture inside the constriction (R1 at $x_f = 2.0$, R2 at $x_f = 1.4$) and were discarded. Two further runs reached their step ceiling with the cell still inside the constriction (R1 at $x_f = 0.8$) and yield no crossing speed, so the mean at such a driving is over the realizations that crossed. Each single-cell simulation initialized one RBC $5~\si{\micro\metre}$ upstream of the constriction entrance, relaxed it for $2 \times 10^{4}$ steps without driving, and then drove it until its center of mass had passed $5~\si{\micro\metre}$ beyond the constriction exit; the speed of Fig~\ref{fig:microchannel}B is the mean speed of the center of mass from $5$ to $20~\si{\micro\metre}$ past the constriction entrance. Periodic boundaries were applied in the flow direction and the walls were no-slip; in this channel the solvent--wall repulsion is stronger than the solvent--solvent one and the four walls of the $10 \times 10~\si{\micro\metre}$ channel also reflect solvent and membrane particles, both to keep the solvent out of the walls (Table~S1 in S1 Text). The walls of the constriction itself are frozen DPD particles without such a reflecting boundary, so membrane particles interact with them, and with one another, through the same purely repulsive Weeks--Chandler--Andersen contact term as the slit assays below ($\epsilon = 1$ in model units, $\sigma = 0.40~\si{\micro\metre}$, cut off at its minimum $2^{1/6}\sigma = 0.449~\si{\micro\metre}$). A contact term of that range excludes the membrane from a wall only where the wall particles lie closer together than its cutoff, so the faces of the constriction carry an additional monolayer of frozen particles on a square lattice of pitch $0.30~\si{\micro\metre}$: every point of such a face is within $0.30/\sqrt{2} = 0.21~\si{\micro\metre}$ of a lattice site, inside the cutoff, so the face cannot be crossed. Each monolayer is set back $0.40~\si{\micro\metre}$ into the solid rather than on the nominal face, so that the contact term excludes the membrane to just outside that face instead of a full cutoff away from it; a layer placed on the face itself would hold the membrane $0.45~\si{\micro\metre}$ clear of each wall and narrow the channel by a third. Membrane configurations saved every $5 \times 10^{3}$ steps were checked against the channel geometry, and no membrane particle entered a wall in any run; the closest any membrane particle came to a constriction face was $0.16~\si{\micro\metre}$ over the 75 audited runs. The aperture available to the membrane is accordingly $4.90 \times 2.60~\si{\micro\metre}$ rather than the nominal $5.0 \times 2.7$: both dimensions are inset by the same $0.098~\si{\micro\metre}$, which is $2.0\%$ of the width and $3.6\%$ of the depth, and the solvent, which its own wall repulsion holds off as well, fills only $\sim\!4.5 \times 2.1~\si{\micro\metre}$ of the nominal cross-section (the extent over which its density exceeds half its bulk value), which lowers the flow at a given driving. The absolute speeds of Fig~\ref{fig:microchannel}B therefore belong to this effective geometry and to the single time conversion of this work (below); the four cell types share both, so the comparison between them is like-for-like.

\textit{Splenic slit traversal.} Inter-endothelial-slit (IES) traversal was modeled using a slit formed between two parallel cylindrical endothelial elements of diameter $2.5~\si{\micro\metre}$, whose axes run across the channel and whose facing surfaces are separated by $1.5~\si{\micro\metre}$. The resulting aperture is $1.5~\si{\micro\metre}$ wide and $5.5~\si{\micro\metre}$ tall, and the cylinder diameter sets its $2.5~\si{\micro\metre}$ streamwise depth; it is cut in a block of frozen particles that spans the cross-section of a $30 \times 12.5 \times 12.5~\si{\micro\metre\cubed}$ box, periodic in all three directions and without side walls, so that the geometry is a periodic array of slits. All geometric values quoted here were measured directly from the simulation input configuration. The frozen block confines the membrane without a reflecting boundary, so the saved configurations were checked against that same geometry for membrane particles inside the endothelium: across the $25$ runs of this assay and of its surface-to-volume test, no membrane particle entered a cylinder or the block that caps the aperture, the closest approach being $0.35~\si{\micro\metre}$ outside a cylinder surface. A single RBC was placed with its center of mass $9~\si{\micro\metre}$ upstream of the plane of the slit (the plane through the cylinder axes), and the flow was driven by a constant trans-slit pressure gradient of $\Delta P / L = 6$~Pa/\si{\micro\metre}, the value applied to this same slit geometry by Liu \textit{et al.}~\cite{LIU20214heated}. This value sits at the lower end of the $5$--$20$~Pa/\si{\micro\metre} applied to splenic-slit geometries by Li \textit{et al.}~\cite{li2021How}, and was adopted to match those computational protocols rather than from an independent calibration against splenic pressures. Because the time scale of this work is fixed by the control transit through this slit (below), the comparison with the in vivo transit times of MacDonald \textit{et al.}~\cite{macdonald1987kinetics} (Fig~\ref{fig:microchannel}D) places the reticulocyte delays within the in vivo distribution but does not test the loading independently. Each cell type was driven from its initial configuration, from the first step, in three independent realizations of $2.2 \times 10^{5}$ steps, and a passage time is the first time at which the center of mass reached the plane of the slit, $9~\si{\micro\metre}$ from its start, or a plane $2.5~\si{\micro\metre}$ further downstream; one R2 realization ended in a numerical membrane rupture after its center of mass had passed the slit plane and was replaced by a fourth. For the surface-to-volume test the reference volume of each model was lowered, at the control membrane area, to $V_0 = 92.46 \times 1.44/(S/V)$~fL, which gives it the $S/V$ of Table~\ref{tab:mechanical_properties}, and the cell was relaxed for $2 \times 10^{4}$ steps without driving before release.

\textit{Pairwise single-file transport.} Leader--follower passage (Fig~\ref{fig:pairwise}) was modeled in a longer channel of $250 \times 11 \times 15~\si{\micro\metre\cubed}$ containing a sub-cellular slit $1.5~\si{\micro\metre}$ wide, $5~\si{\micro\metre}$ tall and $4~\si{\micro\metre}$ deep, with a leader and a trailing cell (the follower) arranged in single file upstream. The two posts forming the slit are periodic in the transverse direction, so each period presents two identical $1.5~\si{\micro\metre}$ apertures---one between the posts and one across the transverse periodic boundary---and the fluid is free to pass through both; the cells were initialized in front of the interior aperture, and every passage went through it, the passing cell's center of mass crossing the post axis at $y = 5.3$--$5.7~\si{\micro\metre}$ against a seam at $y = 0$ and $11~\si{\micro\metre}$; $18$ followers retained at drivings below their thresholds drifted toward the seam aperture without passing. Boundaries were periodic in the flow and transverse directions, with no-slip walls above and below the channel. Every run started from the same configuration, obtained by relaxing the two cells for $4 \times 10^{3}$ steps without driving, with the leader $1.6~\si{\micro\metre}$ from the nearest post; the runs of the follower alone started from it with the leader removed. Besides the soft DPD repulsion, membrane particles interacted through a purely repulsive Weeks--Chandler--Andersen term, $U(r) = 4\epsilon[(\sigma/r)^{12} - (\sigma/r)^{6}] + \epsilon$ for $r < 2^{1/6}\sigma$, with $\epsilon = 1$ in model units ($\approx\!11\,k_BT$) and $\sigma = 0.4~\si{\micro\metre}$, between the two cells, between particles of one membrane more than three bonds apart, and between membrane and post particles; $\sigma$ was raised from $0.2~\si{\micro\metre}$ to its final value over the first $3 \times 10^{3}$ steps. That term acts between particles, which are $0.56~\si{\micro\metre}$ apart on average and up to $1.12~\si{\micro\metre}$ where the membrane is stretched, so it cannot by itself keep the two faces of a tight crease apart. The runs therefore carry a second repulsive term of the same form acting on the distance from each membrane particle to the nearest point of every membrane triangle more than three bonds away, with $\epsilon = 1$ in model units and $\sigma = 0.15~\si{\micro\metre}$, the reaction distributed over the triangle's three vertices by the barycentric weights of that nearest point. Its range is a third of the particle term's, so it is inert in ordinary contact between two membranes, where the particle term already sets the separation, and acts only where a surface folds onto itself. Because a particle-to-triangle distance, unlike a distance between two particles, is not bounded away from zero, the magnitude of this force is capped at $30$ in model units, a value set by the time step so that the velocity it can impart in one step stays below a tenth of the thermal velocity; the work required to cross the surface is still the cap times the range, some $50~k_{B}T$. The follower alone and behind each of the three leaders was driven at $x_f = 0.30$--$1.20$ in steps of $0.05$, with three independent realizations at every driving ($228$ runs of $2.2 \times 10^{5}$ steps, $1215$~ms). Passage was scored as for the single-cell ladder below: the follower's passage time runs from release until its center of mass reaches the slit-exit plane, a run is retained if that time exceeds the $1200$~ms criterion, the critical driving $x_{f,c}$ is the driving at which the realization-averaged passage time crosses the criterion, and its uncertainty is the standard deviation of the per-realization thresholds. For the follower alone, $x_{f,c} = 0.48 \pm 0.02$. The control rung of the single-cell ladder of Fig~\ref{fig:dpc_ladder} is the same configuration carrying the particle contact term alone, for which $x_{f,c} = 0.46 \pm 0.01$; the two agree within their uncertainties, as they should, since the vertex-to-face term acts only where a membrane folds onto itself and a cell driven alone does not. The follower's arrival at the slit was taken as its center of mass reaching $4~\si{\micro\metre}$ upstream of the post axis, its transit as the time from arrival to the exit plane, and the leader's clearance as the leader's center of mass reaching the exit plane. Membrane configurations were saved every $2 \times 10^{3}$ steps and audited for contact failures: in all $25{,}308$ saved configurations no volume lay inside both cells (on a $0.25~\si{\micro\metre}$ grid), no membrane particle entered a post (closest approach $0.38~\si{\micro\metre}$ outside the post surface), and the particles the contact terms act on, whether on different cells or on one membrane, stayed at least $0.38~\si{\micro\metre}$ apart.

\textit{Single-cell critical driving.} The population thresholds of Fig~\ref{fig:dpc_ladder}B were measured in this same elongated slit channel (Fig~\ref{fig:dpc_ladder}A) with a single cell released from the same upstream position in every run, varying only the membrane shear modulus: the CTR discocyte and the R1--R3 reticulocytes at the moduli of Table~\ref{tab:mechanical_properties}, and a sickled cell at $20\times$ and $50\times$ the control modulus ($94.6$ and $236~\si{\micro\newton\per\metre}$), bracketing the mean shear modulus measured for irreversibly sickled cells ($135~\si{\micro\newton\per\metre}$)~\cite{Williams2025Quantifying}. Every run started from a relaxed two-cell configuration of this channel with its leading cell removed. Membrane area, volume and bending modulus were held at the control values throughout, so that the ladder isolates the contribution of the shear modulus; restoring the surface-to-volume ratio of Table~\ref{tab:mechanical_properties} alongside the modulus \emph{lowers} the penalty rather than raising it, in all three subtypes of the inter-endothelial-slit assay (S1 Text, Section~7), so the shear-modulus-only ladder is an upper bound on the reticulocyte penalty. The driving was stepped in increments of $0.05$ in the dimensionless body-force parameter of the assay, with three independent realizations at every driving. The passage time was taken from release to the slit-exit plane, and a cell was scored as retained if that time exceeded the $1200$~ms retention criterion ($2.17 \times 10^{5}$ steps); runs that exhausted their step budget without clearing the slit were treated as retained lower bounds. The posts that form the slit are frozen DPD particles, and membrane particles interact with them, and with one another, through the same purely repulsive Weeks--Chandler--Andersen contact term as the pairwise assay ($\epsilon = 1$ in model units, $\sigma = 0.40~\si{\micro\metre}$, cut off at its minimum $2^{1/6}\sigma$). Because each run starts from a relaxed configuration with the cell already in the slit, $\sigma$ is introduced in four stages ($0.20$, $0.30$, $0.35$ and $0.40~\si{\micro\metre}$). Every saved configuration was then checked against the analytic post surfaces: no membrane particle entered a post in any run of the ladder, including both sickled rungs, the closest approach being $0.16~\si{\micro\metre}$ outside a post surface, and the closest membrane--post particle pair, $0.35~\si{\micro\metre}$, lies inside the cutoff, so the contact term is active and not merely present. The threshold is the driving at which the realization-averaged passage time crosses the criterion, interpolated linearly between the bracketing drivings, and its uncertainty is the standard deviation of the thresholds obtained from the individual realizations. This criterion is the retention time used elsewhere in this work. Re-scoring the same runs within a shorter window ($1.2 \times 10^{5}$ steps, $663$~ms) raises the control threshold from $0.46$ to $0.71$ and lifts the three reticulocyte thresholds above the highest driving simulated, so that window bounds them rather than locating them; the ordering does not depend on where the criterion is placed, however, because the mean passage time already increases with shear modulus at every driving at which two or more of these models are compared (S1 Text, Section~7).

\textit{Shear flow.} The viscosity curves of Fig~\ref{fig:viscosity} were computed in~\cite{chai2025silico}, in a rectangular box of $60 \times 60 \times 50~\si{\micro\metre\cubed}$, bounded by two no-slip walls of thickness $5~\si{\micro\metre}$. Shear flow was imposed by translating the walls in opposite directions at constant velocity, giving imposed shear rates $\dot{\gamma} \in [1, 10^{3}]~\mathrm{s}^{-1}$. Periodic boundaries were applied in the flow ($x$) and vorticity ($z$) directions. Each RBC was represented by 500 membrane particles, and 558 cells were modeled to achieve a hematocrit of $H_t = 36\%$; plasma was represented by $\sim\!6 \times 10^{5}$ DPD beads, yielding a total system size of $\sim\!8.8 \times 10^{5}$ particles.

Across all simulations, the cytoplasm was modeled using DPD fluid particles identical to plasma, while the RBC membrane consisted of bonded DPD particles forming a triangulated network. The simulations of this work ranged from $\sim\!1.8 \times 10^{4}$ particles (single-cell stretching, S1 Text, Section~4) to $\sim\!2.2 \times 10^{5}$ particles (the elongated slit channel of the pairwise and critical-driving assays), and the longest runs lasted $9.3 \times 10^{5}$ DPD time steps. All of them were performed with an extended version of the LAMMPS code at the Center for Computation and Visualization at Brown University. Physical times quoted in this work use a single conversion, $1$ DPD step $= 5.52~\si{\micro\second}$, applied unchanged to every assay. It is fixed by the control-discocyte traversal of the IES slit, which takes $4.53 \times 10^{4}$ steps from release to the exit plane and is reported as $250$~ms (Fig~\ref{fig:microchannel}D); on the same scale the $1200$~ms retention criterion corresponds to $2.17 \times 10^{5}$ steps. We state the conversion explicitly because it is a viscous mapping calibrated on the transit itself, not a scale derived from the DPD length, mass and energy units, and the two do not agree. The values behind every simulated panel, for each realization and as plotted, are listed in \nameref{S1_Data}.


\section*{Supporting information}

\paragraph*{S1 Text.}
\label{S1_Text}
{\bf Supplementary methods, model calibration, and robustness analyses.} Full DPD force definitions, parameters, and unit mappings; the membrane model and the mechanical parameters of the control and reticulocyte models; checks of the membrane model; confined-flow and viscosity protocols; supravital NMB staining by morphological class; sensitivity of the critical-driving ladder to the scoring criterion and to excess area; the flow-induced shape transition with the shear modulus alone.

\paragraph*{S1 Data.}
\label{S1_Data}
{\bf Values behind the figures.} One CSV file per plotted panel of Fig 1B, Figs 4--8 and Figs S1 and S2 in \nameref{S1_Text} (for the simulations, the value from every realization as well as the plotted mean and spread), the wet-mount footprint census of Methods, and the membrane-stretching runs and resting-shape profile of Section~4 of \nameref{S1_Text}; a README file describes each file and its columns.

\section*{Acknowledgments}
Simulations were carried out at the Center for Computation and Visualization of Brown University.

\section*{Data availability}
All data are in the manuscript and/or Supporting information files.

\section*{Competing interests}
The authors have declared that no competing interests exist.

\nolinenumbers

\bibliography{ref_main}

\begin{thebibliography}{10}

\bibitem{Mebius2005Structure}
Mebius RE, Kraal G.
\newblock Structure and function of the spleen.
\newblock Nature Reviews Immunology. 2005;5(8):606--616.
\newblock doi:{10.1038/nri1669}.

\bibitem{Cesta2006Normal}
Cesta MF.
\newblock Normal structure, function, and histology of the spleen.
\newblock Toxicologic Pathology. 2006;34(5):455--465.
\newblock doi:{10.1080/01926230600867743}.

\bibitem{Li2018Mechanics}
Li H, Lu L, Li X, Buffet PA, Dao M, Karniadakis GE, et~al.
\newblock Mechanics of diseased red blood cells in human spleen and
  consequences for hereditary blood disorders.
\newblock Proceedings of the National Academy of Sciences.
  2018;115(38):9574--9579.
\newblock doi:{10.1073/pnas.1806501115}.

\bibitem{moreau2023physical}
Moreau A, Yaya F, Lu H, Surendranath A, Charrier A, Dehapiot B, et~al.
\newblock Physical mechanisms of red blood cell splenic filtration.
\newblock Proceedings of the National Academy of Sciences.
  2023;120(44):e2300095120.
\newblock doi:{10.1073/pnas.2300095120}.

\bibitem{evans1973new}
Evans EA.
\newblock New membrane concept applied to the analysis of fluid shear- and
  micropipette-deformed red blood cells.
\newblock Biophysical Journal. 1973;13(9):941--954.
\newblock doi:{10.1016/s0006-3495(73)86036-9}.

\bibitem{dao2021erythrocyte}
Dao M, MacDonald I, Asaro RJ.
\newblock Erythrocyte flow through the interendothelial slits of the splenic
  venous sinus.
\newblock Biomechanics and Modeling in Mechanobiology. 2021;20(6):2227--2245.
\newblock doi:{10.1007/s10237-021-01503-y}.

\bibitem{mohandas2008red}
Mohandas N, Gallagher PG.
\newblock Red cell membrane: past, present, and future.
\newblock Blood. 2008;112(10):3939--3948.
\newblock doi:{10.1182/blood-2008-07-161166}.

\bibitem{baskurt2003blood}
Baskurt OK, Meiselman HJ.
\newblock Blood rheology and hemodynamics.
\newblock Seminars in Thrombosis and Hemostasis. 2003;29:435--450.
\newblock doi:{10.1055/s-2003-44551}.

\bibitem{nader2019blood}
Nader E, Skinner S, Romana M, Fort R, Lemonne N, Guillot N, et~al.
\newblock Blood rheology: key parameters, impact on blood flow, role in sickle
  cell disease and effects of exercise.
\newblock Frontiers in Physiology. 2019;10:1329.
\newblock doi:{10.3389/fphys.2019.01329}.

\bibitem{perazzo2022effect}
Perazzo A, Peng Z, Young YN, Feng Z, Wood DK, Higgins JM, et~al.
\newblock The effect of rigid cells on blood viscosity: linking rheology and
  sickle cell anemia.
\newblock Soft Matter. 2022;18(3):554--565.
\newblock doi:{10.1039/d1sm01299a}.

\bibitem{chasis1989membrane}
Chasis JA, Prenant M, Leung A, Mohandas N.
\newblock Membrane assembly and remodeling during reticulocyte maturation.
\newblock Blood. 1989;74(3):1112--1120.
\newblock doi:{10.1182/blood.v74.3.1112.1112}.

\bibitem{clark1988senescence}
Clark MR.
\newblock Senescence of red blood cells: progress and problems.
\newblock Physiological Reviews. 1988;68(2):503--554.
\newblock doi:{10.1152/physrev.1988.68.2.503}.

\bibitem{malleret2013significant}
Malleret B, Xu F, Mohandas N, Suwanarusk R, Chu C, Leite JA, et~al.
\newblock Significant biochemical, biophysical and metabolic diversity in
  circulating human cord blood reticulocytes.
\newblock PLoS ONE. 2013;8(10):e76062.
\newblock doi:{10.1371/journal.pone.0076062}.

\bibitem{Renoux2019Impact}
Renoux C, Faivre M, Bessaa A, Da~Costa L, Joly P, Gauthier A, et~al.
\newblock Impact of surface-area-to-volume ratio, internal viscosity and
  membrane viscoelasticity on red blood cell deformability measured in isotonic
  condition.
\newblock Scientific Reports. 2019;9:6771.
\newblock doi:{10.1038/s41598-019-43200-y}.

\bibitem{mel1977reticulocyte}
Mel HC, Prenant M, Mohandas N.
\newblock Reticulocyte motility and form: studies on maturation and
  classification.
\newblock Blood. 1977;49(6):1001--1009.
\newblock doi:{10.1182/blood.v49.6.1001.bloodjournal4961001}.

\bibitem{gronowicz1984maturation}
Gronowicz G, Swift H, Steck TL.
\newblock Maturation of the reticulocyte in vitro.
\newblock Journal of Cell Science. 1984;71(1):177--197.

\bibitem{StevensHernandez2022Reticulocyte}
Stevens-Hernandez CJ, Flatt JF, Kupzig S, Bruce LJ.
\newblock Reticulocyte maturation and variant red blood cells.
\newblock Frontiers in Physiology. 2022;13:834463.
\newblock doi:{10.3389/fphys.2022.834463}.

\bibitem{Leduc2025Comprehensive}
Leduc M, Papoin J, Yan H, Le~Gall M, El~Nemer W, Hillyer C, et~al.
\newblock Comprehensive phenotypic and proteomic analyses of human reticulocyte
  maturation.
\newblock Blood Red Cells \& Iron. 2025;1(2):100012.
\newblock doi:{10.1016/j.brci.2025.100012}.

\bibitem{pivkin2016biomechanics}
Pivkin IV, Peng Z, Karniadakis GE, Buffet PA, Dao M, Suresh S.
\newblock Biomechanics of red blood cells in human spleen and consequences for
  physiology and disease.
\newblock Proceedings of the National Academy of Sciences.
  2016;113(28):7804--7809.
\newblock doi:{10.1073/pnas.1606751113}.

\bibitem{li2021How}
Li H, Liu ZL, Lu L, Buffet P, Karniadakis GE.
\newblock How the spleen reshapes and retains young and old red blood cells: a
  computational investigation.
\newblock PLoS Computational Biology. 2021;17(11):e1009516.
\newblock doi:{10.1371/journal.pcbi.1009516}.

\bibitem{macdonald1987kinetics}
MacDonald IC, Ragan DM, Schmidt EE, Groom AC.
\newblock Kinetics of red blood cell passage through interendothelial slits
  into venous sinuses in rat spleen, analyzed by in vivo microscopy.
\newblock Microvascular Research. 1987;33(1):118--134.
\newblock doi:{10.1016/0026-2862(87)90011-2}.

\bibitem{safeukui2012quantitative}
Safeukui I, Buffet PA, Deplaine G, Perrot S, Brousse V, Ndour A, et~al.
\newblock Quantitative assessment of sensing and sequestration of spherocytic
  erythrocytes by the human spleen.
\newblock Blood. 2012;120(2):424--430.
\newblock doi:{10.1182/blood-2012-01-404103}.

\bibitem{safeukui2018sensing}
Safeukui I, Buffet PA, Deplaine G, Perrot S, Brousse V, Sauvanet A, et~al.
\newblock Sensing of red blood cells with decreased membrane deformability by
  the human spleen.
\newblock Blood Advances. 2018;2(20):2581--2587.
\newblock doi:{10.1182/bloodadvances.2018024562}.

\bibitem{chai2026multiscale}
Chai Z, Ahmadi~Daryakenari N, Karniadakis GE.
\newblock A multiscale signaling--biophysical framework reveals mechanisms of
  macrophage-mediated {RBC} clearance in sickle cell and {G}aucher disease.
\newblock PNAS Nexus. 2026;pgag348.
\newblock doi:{10.1093/pnasnexus/pgag348}.

\bibitem{stewart2002spleen}
Stewart IB, McKenzie DC.
\newblock The Human Spleen During Physiological Stress.
\newblock Sports Medicine. 2002;32(6):361--369.
\newblock doi:{10.2165/00007256-200232060-00002}.

\bibitem{roach2018lakelouise}
Roach RC, Hackett PH, Oelz O, B{\"a}rtsch P, Luks AM, MacInnis MJ, et~al.
\newblock The 2018 {L}ake {L}ouise Acute Mountain Sickness Score.
\newblock High Altitude Medicine \& Biology. 2018;19(1):4--6.
\newblock doi:{10.1089/ham.2017.0164}.

\bibitem{stauffer2024making}
Stauffer E, Pichon AP, Champigneulle B, Furian M, Hancco I, Darras A, et~al.
\newblock Making a virtue out of an evil: are red blood cells from chronic
  mountain sickness patients eligible for transfusions?
\newblock American Journal of Hematology. 2024;99(7):1407--1410.
\newblock doi:{10.1002/ajh.27317}.

\bibitem{goodman2014splenic}
Goodman J, Hassell K, Irwin D, Witkowski EH, Nuss R.
\newblock The Splenic Syndrome in Individuals with Sickle Cell Trait.
\newblock High Altitude Medicine \& Biology. 2014;15(4):468--471.
\newblock doi:{10.1089/ham.2014.1034}.

\bibitem{quinn2011combined}
Quinn DJ, Pivkin I, Wong SY, Chiam KH, Dao M, Karniadakis GE, et~al.
\newblock Combined simulation and experimental study of large deformation of
  red blood cells in microfluidic systems.
\newblock Annals of Biomedical Engineering. 2011;39(3):1041--1050.
\newblock doi:{10.1007/s10439-010-0232-y}.

\bibitem{qiang2023microfluidic}
Qiang Y, Sissoko A, Liu ZL, Dong T, Zheng F, Kong F, et~al.
\newblock Microfluidic study of retention and elimination of abnormal red blood
  cells by human spleen with implications for sickle cell disease.
\newblock Proceedings of the National Academy of Sciences.
  2023;120(6):e2217607120.
\newblock doi:{10.1073/pnas.2217607120}.

\bibitem{recktenwald2024morphology}
Recktenwald SM, Rashidi Y, Graham I, Arratia PE, Del~Giudice F, Wagner C.
\newblock Morphology, repulsion, and ordering of red blood cells in
  viscoelastic flows under confinement.
\newblock Soft Matter. 2024;20(25):4950--4963.
\newblock doi:{10.1039/d4sm00446a}.

\bibitem{tang2017openrbc}
Tang YH, Lu L, Li H, Evangelinos C, Grinberg L, Sachdeva V, et~al.
\newblock {OpenRBC}: a fast simulator of red blood cells at protein resolution.
\newblock Biophysical Journal. 2017;112(10):2030--2037.
\newblock doi:{10.1016/j.bpj.2017.04.020}.

\bibitem{chai2023dynamics}
Chai Z, Gu S, Lykotrafitis G.
\newblock Dynamics of the axon plasma membrane skeleton.
\newblock Soft Matter. 2023;19(14):2514--2528.
\newblock doi:{10.1039/d2sm01602h}.

\bibitem{chai2022periodic}
Chai Z, Tzingounis AV, Lykotrafitis G.
\newblock The periodic axon membrane skeleton leads to {N}a nanodomains but
  does not impact action potentials.
\newblock Biophysical Journal. 2022;121(18):3334--3344.
\newblock doi:{10.1016/j.bpj.2022.08.027}.

\bibitem{chang2016md}
Chang HY, Li X, Li H, Karniadakis GE.
\newblock {MD/DPD} multiscale framework for predicting morphology and stresses
  of red blood cells in health and disease.
\newblock PLOS Computational Biology. 2016;12(10):e1005173.
\newblock doi:{10.1371/journal.pcbi.1005173}.

\bibitem{zhang2021deep}
Zhang Y, Chai Z, Lykotrafitis G.
\newblock Deep reinforcement learning with a particle dynamics environment
  applied to emergency evacuation of a room with obstacles.
\newblock Physica A: Statistical Mechanics and its Applications.
  2021;571:125845.
\newblock doi:{10.1016/j.physa.2021.125845}.

\bibitem{zhang2020deep}
Zhang Y, Chai Z, Sun Y, Lykotrafitis G.
\newblock A deep reinforcement learning model based on deterministic policy
  gradient for collective neural crest cell migration.
\newblock arXiv preprint arXiv:2007{\char46}03190. 2020.

\bibitem{fedosov2010multiscale}
Fedosov DA, Caswell B, Karniadakis GE.
\newblock A multiscale red blood cell model with accurate mechanics, rheology,
  and dynamics.
\newblock Biophysical Journal. 2010;98(10):2215--2225.
\newblock doi:{10.1016/j.bpj.2010.02.002}.

\bibitem{Fedosov2011Multiscale}
Fedosov DA, Lei H, Caswell B, Suresh S, Karniadakis GE.
\newblock Multiscale modeling of red blood cell mechanics and blood flow in
  malaria.
\newblock PLOS Computational Biology. 2011;7(12):e1002270.
\newblock doi:{10.1371/journal.pcbi.1002270}.

\bibitem{Fedosov2011Predicting}
Fedosov DA, Pan W, Caswell B, Gompper G, Karniadakis GE.
\newblock Predicting human blood viscosity in silico.
\newblock Proceedings of the National Academy of Sciences.
  2011;108(29):11772--11777.
\newblock doi:{10.1073/pnas.1101210108}.

\bibitem{groot1997dissipative}
Groot RD, Warren PB.
\newblock Dissipative particle dynamics: bridging the gap between atomistic and
  mesoscopic simulation.
\newblock The Journal of Chemical Physics. 1997;107(11):4423--4435.
\newblock doi:{10.1063/1.474784}.

\bibitem{hoogerbrugge1992simulating}
Hoogerbrugge PJ, Koelman JMVA.
\newblock Simulating microscopic hydrodynamic phenomena with dissipative
  particle dynamics.
\newblock Europhysics Letters. 1992;19(3):155--160.
\newblock doi:{10.1209/0295-5075/19/3/001}.

\bibitem{chai2025silico}
Chai Z, Li G, Ndour PA, Connes P, Buffet PA, Franco M, et~al.
\newblock In silico biophysics and rheology of blood and red blood cells in
  {G}aucher disease.
\newblock PLOS Computational Biology. 2025;21(9):e1012705.
\newblock doi:{10.1371/journal.pcbi.1012705}.

\bibitem{cluitmans2012red}
Cluitmans JCA, Hardeman MR, Dinkla S, Brock R, Bosman GJCGM.
\newblock Red blood cell deformability during storage: towards functional
  proteomics and metabolomics in the blood bank.
\newblock Blood Transfusion. 2012;10(Suppl 2):s12--s18.
\newblock doi:{10.2450/2012.004S}.

\bibitem{chai2026quantifying}
Chai Z, Zheng J, Li H, Dao M, Karniadakis GE.
\newblock Quantifying the biophysical properties of stomatocytes in health and
  disease.
\newblock Biophysical Journal. 2026;in press.
\newblock doi:{10.1016/j.bpj.2026.10.007}.

\bibitem{chai2026pocked}
Chai Z, Zheng J, Li H, Dao M, Karniadakis GE.
\newblock The typical red cell makes too little vacuole material for
  coalescence to build the large vacuoles of pocked cells.
\newblock arXiv preprint arXiv:2610{\char46}22646. 2026.

\bibitem{siebenmann2017regulation}
Siebenmann C, Robach P, Lundby C.
\newblock Regulation of Blood Volume in Lowlanders Exposed to High Altitude.
\newblock Journal of Applied Physiology. 2017;123(4):957--966.
\newblock doi:{10.1152/japplphysiol.00118.2017}.

\bibitem{gaur2021reticulocyte}
Gaur M, Sehgal T.
\newblock Reticulocyte count: a simple test but tricky interpretation!
\newblock Pan African Medical Journal. 2021;40:3.
\newblock doi:{10.11604/pamj.2021.40.3.31316}.

\bibitem{suresh2005connections}
Suresh S, Spatz J, Mills JP, Micoulet A, Dao M, Lim CT, et~al.
\newblock Connections between single-cell biomechanics and human disease
  states: gastrointestinal cancer and malaria.
\newblock Acta Biomaterialia. 2005;1(1):15--30.
\newblock doi:{10.1016/j.actbio.2004.09.001}.

\bibitem{fedosov2014deformation}
Fedosov DA, Peltom{\"a}ki M, Gompper G.
\newblock Deformation and dynamics of red blood cells in flow through
  cylindrical microchannels.
\newblock Soft Matter. 2014;10(24):4258--4267.
\newblock doi:{10.1039/c4sm00248b}.

\bibitem{LIU20214heated}
Liu ZL, Li H, Qiang Y, Buffet P, Dao M, Karniadakis GE.
\newblock Computational modeling of biomechanics and biorheology of heated red
  blood cells.
\newblock Biophysical Journal. 2021;120(21):4663--4671.
\newblock doi:{10.1016/j.bpj.2021.09.038}.

\bibitem{chai2026gaucher}
Chai Z, de~Person M, Buffet PA, Franco M, Karniadakis GE.
\newblock Quantifying the biophysical properties of red blood cells in
  {G}aucher disease.
\newblock Biophysical Journal. 2026;in press.

\bibitem{Williams2025Quantifying}
Williams DC, Higgins JM, Wood DK.
\newblock Quantifying the unique mechanical properties of irreversibly sickled
  cells in sickle cell disease.
\newblock Blood Vessels, Thrombosis \& Haemostasis. 2025;2(3):100077.
\newblock doi:{10.1016/j.bvth.2025.100077}.

\bibitem{Xu2016Stiffening}
Xu Z, Zheng Y, Wang X, Shehata N, Wang C, Xie S, et~al.
\newblock Stiffening of sickle cell trait red blood cells under simulated
  strenuous exercise conditions.
\newblock Microsystems \& Nanoengineering. 2016;2:16061.
\newblock doi:{10.1038/micronano.2016.61}.

\bibitem{Chien1970Shear}
Chien S.
\newblock Shear dependence of effective cell volume as a determinant of blood
  viscosity.
\newblock Science. 1970;168(3934):977--979.
\newblock doi:{10.1126/science.168.3934.977}.

\bibitem{SKOVBORG1966129}
Skovborg F, Nielsen AV, Schlichtkrull J, Ditzel J.
\newblock Blood-viscosity in diabetic patients.
\newblock The Lancet. 1966;287(7429):129--131.
\newblock doi:{10.1016/s0140-6736(66)91264-5}.

\bibitem{franco2013abnormal}
Franco M, Collec E, Connes P, van~den Akker E, Billette~de Villemeur T,
  Belmatoug N, et~al.
\newblock Abnormal properties of red blood cells suggest a role in the
  pathophysiology of {G}aucher disease.
\newblock Blood. 2013;121(3):546--555.
\newblock doi:{10.1182/blood-2012-07-442467}.

\bibitem{maron1956application}
Maron SH, Pierce PE.
\newblock Application of {R}ee--{E}yring generalized flow theory to suspensions
  of spherical particles.
\newblock Journal of Colloid Science. 1956;11(1):80--95.
\newblock doi:{10.1016/0095-8522(56)90023-x}.

\bibitem{JAVADI2021hyperviscosity}
Javadi E, Deng Y, Karniadakis GE, Jamali S.
\newblock In silico biophysics and hemorheology of blood hyperviscosity
  syndrome.
\newblock Biophysical Journal. 2021;120(13):2723--2733.
\newblock doi:{10.1016/j.bpj.2021.05.013}.

\bibitem{javadi2022circulating}
Javadi E, Li H, Gallastegi AD, Frydman GH, Jamali S, Karniadakis GE.
\newblock Circulating cell clusters aggravate the hemorheological abnormalities
  in COVID-19.
\newblock Biophysical Journal. 2022;121(18):3309--3319.
\newblock doi:{10.1016/j.bpj.2022.08.031}.

\bibitem{baskurt2009guidelines}
Baskurt O, Boynard M, Cokelet G, Connes P, Cooke BM, Forconi S, et~al.
\newblock New guidelines for hemorheological laboratory techniques.
\newblock Clinical Hemorheology and Microcirculation. 2009;42(2):75--97.
\newblock doi:{10.3233/ch-2009-1202}.

\bibitem{villafuerte2022highaltitude}
Villafuerte FC, Simonson TS, Bermudez D, Le{\'o}n-Velarde F.
\newblock High-Altitude Erythrocytosis: Mechanisms of Adaptive and Maladaptive
  Responses.
\newblock Physiology. 2022;37(4):175--186.
\newblock doi:{10.1152/physiol.00029.2021}.

\bibitem{storz2016hemoglobin}
Storz JF.
\newblock Hemoglobin-Oxygen Affinity in High-Altitude Vertebrates: Is There
  Evidence for an Adaptive Trend?
\newblock Journal of Experimental Biology. 2016;219(20):3190--3203.
\newblock doi:{10.1242/jeb.127134}.

\bibitem{stembridge2019plasma}
Stembridge M, Williams AM, Gasho C, Dawkins TG, Drane A, Villafuerte FC, et~al.
\newblock The Overlooked Significance of Plasma Volume for Successful
  Adaptation to High Altitude in {S}herpa and {A}ndean Natives.
\newblock Proceedings of the National Academy of Sciences.
  2019;116(33):16177--16179.
\newblock doi:{10.1073/pnas.1909002116}.

\bibitem{beall2007two}
Beall CM.
\newblock Two Routes to Functional Adaptation: {T}ibetan and {A}ndean
  High-Altitude Natives.
\newblock Proceedings of the National Academy of Sciences. 2007;104(suppl
  1):8655--8660.
\newblock doi:{10.1073/pnas.0701985104}.

\bibitem{simonson2010genetic}
Simonson TS, Yang Y, Huff CD, Yun H, Qin G, Witherspoon DJ, et~al.
\newblock Genetic Evidence for High-Altitude Adaptation in {T}ibet.
\newblock Science. 2010;329(5987):72--75.
\newblock doi:{10.1126/science.1189406}.

\bibitem{li2014erythrocyte}
Li H, Lykotrafitis G.
\newblock Erythrocyte membrane model with explicit description of the lipid
  bilayer and the spectrin network.
\newblock Biophysical Journal. 2014;107(3):642--653.
\newblock doi:{10.1016/j.bpj.2014.06.031}.

\bibitem{wang2024two}
Wang S, Ma S, Li H, Dao M, Li X, Karniadakis GE.
\newblock Two-component macrophage model for active phagocytosis with pseudopod
  formation.
\newblock Biophysical Journal. 2024;123(9):1069--1084.
\newblock doi:{10.1016/j.bpj.2024.03.026}.

\bibitem{kumari2024measuring}
Kumari S, Mehendale N, Roy T, Sen S, Mitra D, Paul D.
\newblock Measuring red blood cell deformability and its heterogeneity using a
  fast microfluidic device.
\newblock Cell Reports Physical Science. 2024;5(8):102052.
\newblock doi:{10.1016/j.xcrp.2024.102052}.

\bibitem{man2020microfluidic}
Man Y, Kucukal E, An R, Watson QD, Bosch J, Zimmerman PA, et~al.
\newblock Microfluidic assessment of red blood cell mediated microvascular
  occlusion.
\newblock Lab on a Chip. 2020;20(12):2086--2099.
\newblock doi:{10.1039/d0lc00112k}.

\bibitem{toscano2026graftathena}
Toscano JD, Chai Z, Karniadakis GE.
\newblock {GRAFT-ATHENA}: Self-Improving Agentic Teams for Autonomous Discovery
  and Evolutionary Numerical Algorithms.
\newblock arXiv preprint arXiv:2605{\char46}11117. 2026.

\bibitem{li2016modeling}
Li H, Zhang Y, Ha V, Lykotrafitis G.
\newblock Modeling of band-3 protein diffusion in the normal and defective red
  blood cell membrane.
\newblock Soft matter. 2016;12(15):3643--3653.
\newblock doi:{10.1039/c4sm02201g}.

\bibitem{skalak1973strain}
Skalak R, Tozeren A, Zarda RP, Chien S.
\newblock Strain Energy Function of Red Blood Cell Membranes.
\newblock Biophysical Journal. 1973;13(3):245--264.
\newblock doi:{10.1016/S0006-3495(73)85983-1}.

\bibitem{helfrich1973elastic}
Helfrich W.
\newblock Elastic Properties of Lipid Bilayers: Theory and Possible
  Experiments.
\newblock Zeitschrift f{\"u}r Naturforschung C. 1973;28(11--12):693--703.
\newblock doi:{10.1515/znc-1973-11-1209}.

\bibitem{du2015kinetics}
Du E, Diez-Silva M, Kato GJ, Dao M, Suresh S.
\newblock Kinetics of sickle cell biorheology and implications for painful
  vasoocclusive crisis.
\newblock Proceedings of the National Academy of Sciences.
  2015;112(5):1422--1427.
\newblock doi:{10.1073/pnas.1424111112}.

\bibitem{qiang2019mechanical}
Qiang Y, Liu J, Dao M, Suresh S, Du E.
\newblock Mechanical fatigue of human red blood cells.
\newblock Proceedings of the National Academy of Sciences.
  2019;116(40):19828--19834.
\newblock doi:{10.1073/pnas.1910336116}.

\bibitem{brecher1949new}
Brecher G.
\newblock New methylene blue as a reticulocyte stain.
\newblock American Journal of Clinical Pathology. 1949;19(9):895--896.
\newblock doi:{10.1093/ajcp/19.9\_ts.895}.

\end{thebibliography}

\end{document}